# Communication Complexity and Secure Function Evaluation*

Moni Naor† Kobbi Nissim
Department of Computer Science and Applied Mathematics
Weizmann Institute of Science, Rehovot 76100, Israel
{naor, kobbi}@wisdom.weizmann.ac.il

October 30, 2018

**Abstract**

A secure function evaluation protocol allows two parties to jointly compute a function $f(x, y)$ of their inputs in a manner not leaking more information than necessary. A major result in this field is: "any function $f$ that can be computed using polynomial resources can be computed *securely* using polynomial resources" (where 'resources' refers to communication and computation). This result follows by a general transformation from any circuit for $f$ to a secure protocol that evaluates $f$. Although the resources used by protocols resulting from this transformation are polynomial in the circuit size, they are much higher (in general) than those required for an insecure computation of $f$.

For the design of efficient secure protocols we suggest two new methodologies, that differ with respect to their underlying computational models. In one methodology we utilize the *communication complexity tree* (or branching program) representation of $f$. We start with an efficient (insecure) protocol for $f$ and transform it into a secure protocol. In other words, "any function $f$ that can be computed using communication complexity $c$ can be can be computed securely using communication complexity that is polynomial in $c$ and a security parameter". The second methodology uses the circuit computing $f$, enhanced with *look-up tables* as its underlying computational model. It is possible to simulate any RAM machine in this model with polylogarithmic blowup. Hence it is possible to start with a computation of $f$ on a RAM machine and transform it into a secure protocol.

We show many applications of these new methodologies resulting in protocols efficient either in communication or in computation. In particular, we exemplify a protocol for the "millionaires problem", where two participants want to compare their values but reveal no other information. Our protocol is more efficient than previously known ones in either communication or computation.

†Work done while the author was visiting Stanford University and the IBM Almaden research center.

# 1 Introduction

Two parties, Alice with input $x$ and Bob with input $y$, wish to evaluate a function $z = f(x, y)$ on their joint inputs in a manner that does not allow the other party to gain more information than necessary about their inputs, i.e. more than can be deduced from $z$. A protocol that allows Alice and Bob to achieve this is known as *secure function evaluation* of $f$. This problem generalizes many cryptographic tasks and has been extensively investigated in the last twenty years[1]. One of the most important results of the area is that "anything that can computed efficiently can be securely evaluated in an efficient manner". More precisely, if the circuit complexity of $f(x, y)$ is $s$, then there is a secure function evaluation protocol operating in time polynomial in $s$ and a security parameter, provided certain cryptographic primitives exist.

The main goal of this work is to provide the communication complexity counterpart of this result. Informally, if $f(x, y)$ has communication complexity $c$, then there is a secure function evaluation protocol for $f$ with communication complexity which is polynomial in $c$ and a security parameter, provided certain cryptographic primitives exist.

Being polynomial in the communication complexity (as compared to the circuit complexity) of a function leads to more efficient protocols for functions over large data held by the parties. For many such functions, the communication complexity is significantly lower than the size of the best known circuit. In particular, the communication complexity may be *sublinear* in the input size. As an example consider computing the median: Alice and Bob each hold subsets $x, y$ of $\{1, \ldots, n\}$ and they wish to compute the median of $x \cup y$ as a multi set. For the median function there exists a protocol with logarithmic (in $n$) communication complexity, hence the potential for a very efficient secure function evaluation protocol.

## 1.1 Related work

Secure function evaluation has been a very active research area and it has been investigated in a variety of models regarding the participating parties behavior, power and number. For any polynomially computable function $f(x_1, x_2, \ldots, x_p)$ it is possible *in principle* to construct a protocol that allows a group of $p$ parties, where party $i$ has as its private input $x_i$, to jointly evaluate $f(x_1, x_2, \ldots, x_p)$ [23]. Following the protocol, the parties learn $f(x_1, x_2, \ldots, x_p)$ but no party $i$ can learn about the other inputs $\{x_j\}_{j \neq i}$ more than can be computed from her own input $x_i$ and the outcome $f(x_1, x_2, \ldots x_p)$. The drawback, however, is that many such protocols are rather complex and require significant interaction between the parties – they do not necessarily correspond to the minimum communication complexity protocol for $f$.

Many of the works in the field followed the 'garbled circuit' construction introduced by Yao [49, 50] and [23] and thus concentrated on the representation of $f$ by a Boolean (combinatorial) circuit. The drawback of this representation is that operations such as accessing a single element from a table result in a relatively large penalty in the circuit size. More specifically, if one follows the 'garbled circuit' construction, as discussed in [43], then the resulting complexity is that of invoking one $OT_1^2$ for each input bit[2] plus a constant number of pseudo-random function evaluations per each gate[3].

[1] One of the first problems considered is Yao's "millionaires problem" [49], where two participants want to check which one has a larger value but leak no other information. Note that the name of the problem has since then been updated to the "*billionaires* problem" ...

**Sublinear communication secure function evaluation protocols** Our work is motivated in part by the attention algorithms for massive datasets have received recently. For many of these problems there are algorithms resulting in communication complexity which is much smaller than the data size, but with no privacy guarantees. Recent results demonstrated that it is possible to create private protocols for some of these tasks, without resorting to the garbled circuit transformation, resulting in protocols with much lower communication complexity, see [16, 17] and [38]. As a consequence of our work, many more of these algorithms can be evaluated securely with low communication overhead.

**Impact of the representation of $f$ on the complexity of secure function evaluation** Researches in the field of secure function evaluation have realized for quite some time that choosing the underlying model is significant when designing a secure function evaluation protocol for a function $f$. In particular, the specific representation chosen for $f$ can have a great impact on the complexity of its secure function evaluation protocol. Some representations of $f$, other than Boolean circuits were used in previous works. To name a few examples, Kilian [28] as well as [23, 5] used permutation branching programs, appealing to Barrington's Theorem. In the multiparty computation with an honest majority setting, Ben-Or et el. [9], followed by many of the works in the area, used a representation of $f$ by an algebraic circuit. Feige et al. [14] and [26] used a representation of $f$ as a product of matrices over a large enough field. Beaver et al. [8] used the representation of $f$ as a low degree polynomial. A recent work by Ishai and Kushilevitz [27] introduced a representation of functions via *randomizing polynomials* and used it to construct round-efficient secure multiparty protocols.

**Communication complexity and privacy** The question of whether it is possible to preserve the communication complexity when constructing protocols for secure function evaluation has been previously considered in the information theoretic setting by Kushilevitz[4]. In the two party case only very special functions have an information theoretic *private* protocol (see characterization in [32] and Kilian [29]). For functions that have private protocols in the information theoretic setting, Kushilevitz [32] showed that their secure version protocol may be much more expensive than their insecure version. There are functions for which any secure evaluation protocol results in exponential communication (whereas linear communication is sufficient for evaluating any function non-privately).

In the computational setting, assuming that oblivious transfer is possible, it turns out that things are quite different: first, as is known from the garbled circuit results [49, 50], it is possible to compute any function privately. What this work shows is, essentially, that the disparity in the communication complexity does not exist: it is possible to compute any function privately while preserving the communication.

[2] $OT_1^2$ is 1-out-of-2 oblivious transfer (see Section 2.2.1) and is a basic building of these protocols.

[3] The protocol described in [20] results in invoking one OT for each gate as well as each input bit.

[4] Kushilevitz addresses the problem of *private* two-party computation of a function. A more general problem of the *secure* computation of a function in a *multiparty* setting with *faulty processors* was addressed by Franklin and Yung [40].

**Circuit complexity vs. Turing Machines and RAM machines** When considering circuits vs. Turing Machines, it is known that there is no significant advantage to the latter. For a Turing Machine $\mathcal{M}$ running in time $T(n)$ there exists a series of circuits $\{\mathcal{C}_n\}_{n\in\mathbb{N}}$ of size $|\mathcal{C}_n| = T(n) \cdot \text{polylog}(T(n))$ computing $f_{\mathcal{M}}$. This result follows from the fact that any computation on a Turing Machine $\mathcal{T}$ may be simulated by an oblivious Turing Machine – where the head position, as a function of time, is independent of the input. The oblivious simulation results in a polylogarithmic blow-up [45]. (See further discussion in Section 5.1.1.)

The situation regarding circuits vs. RAM machines is not known. Since a RAM machine $\mathcal{M}$ running in time $T(n)$ using space $S(n)$ may be simulated by a Turing Machine in time $O(T(n) \cdot S(n))$ (see [48] for more details), we have that it is possible to have a series of circuits of size roughly $O(T(n) \cdot S(n))$ computing $f_{\mathcal{M}}$. It is not known whether it is possible to improve this bound, hence there is a potential gap – a computation on a RAM machine may be much more efficient than any circuit family.

The relationship between circuit complexity and time complexity on a RAM machine has relevance to secure function evaluation protocols – the complexity of many protocols (in particular, those resulting from the garbled circuit transformation) is proportional to the size of the best known circuit for a function. These protocols are rather inefficient when compared with the insecure evaluation of the same function (naturally, on a RAM machine). In part, this is due to lack of an efficient transformation from RAM machines to circuits.

In Section 5 we consider the computational model of *circuits with look-up tables (LUT)* and show how to evaluate it securely and efficiently. We show that for circuits with LUT the gap between the circuit size and a computation on a RAM machine is closed: for a RAM machine $\mathcal{M}$ running in time $T(n)$ using space $S(n)$ there exists a series of circuits with LUT $\{\mathcal{C}_n\}_{n\in\mathbb{N}}$ of size $|\mathcal{C}_n| = T(n) \cdot \text{polylog}(S(n))$ computing $f_{\mathcal{M}}$, where the LUTs are of size $O(S(n))$. These circuits lead, potentially, to much more efficient secure function evaluation protocols.

## 1.2 Methodologies for designing sublinear communication secure protocols

We give two new methodologies for designing secure protocols for a function $f$. In both methodologies, the parties decompose the computation of $f$ so that (i) every party first computes some function of her own input, and (ii) the parties utilize the computed values in an efficient protocol. In other words, suppose $f(x, y)$ can be expressed as $h(g_A(x), g_B(y))$ where $g_A$ and $g_B$ are (efficiently computable) arbitrary functions. If $h$ can be computed securely using sublinear communication, then we get a sublinear secure function evaluation protocol for $f$. The difference between the two methodologies is in the underlying computational model.

**Communication complexity tree** The computational model underlying the first methodology is the communication complexity tree (or branching program) computing $f$. In this model, the functions $g_A, g_B$ correspond to the choices made by the parties during the protocol (as discussed in Section 4, these choices are fully defined by the individual inputs of the parties). The function $h$ corresponds to evaluating the communication complexity tree (or branching program) with respect to the parties' choices.

For many useful and interesting functions the communication complexity of $f$ is *significantly smaller* than its input size. Examples for such functions include the median function (mentioned above), the millionaires problem and Karchmer-Wigderson games.

We propose to utilize the representation of $f$ by its communication complexity tree (or branching program) and show that it can lead to tremendous savings in the *communication* complexity as well as the *computational* complexity of secure function evaluation. For a function $f$ with input size that is relatively large when compared with its communication complexity it is possible to break the cost barrier of an OT invocation per input introduced by the 'garbled circuit' construction.

The general methodology we suggest is to start with a communication efficient (insecure) protocol for the problem and transform it to obtain an efficient secure evaluation; the efficiency can be either in communication or in computation.

As a direct application of this transformation we get an efficient protocol for the secure computation of the median of two subsets $x, y \in \{1, \ldots, n\}$. The communication complexity of this function is $O(\log n)$, and the communication of the resulting secure protocol is polylogarithmic in $n$ times a security parameter.

**Circuits with look-up tables** The second methodology we suggest uses as its underlying computational model the circuit for computing $f$, enhanced with look-up tables. The input wires to a table define an index and the output equals the value stored in the indexed position. This model is equivalent to a write-oblivious RAM (random access memory) machine, where the RAM machine may perform any read operation, but write operations are limited to be oblivious, in the sense that their time and location does not depend on the input. (In other words indirection is allowed with read operations but not with write operations. See discussion in Section 5.)

For a function $f$ whose computation requires a relatively small number of accesses to tables, the size of a circuit with tables for $f$ may be much smaller than that of a Boolean circuit, hence the resulting communication costs of the secure function evaluation protocol may be sublinear in the input size.

Moreover, we show how to simulate RAM (and in particular non-oblivious RAM machines) via circuits with look-up tables with a moderate cost. This allows the design of a (insecure) protocol for computing $f$ on a RAM machine, using a high-level programming language, and then compiling it into a secure protocol.

Again, if $f(x, y)$ can be expressed as $h(g_A(x), g_B(y))$ where $g_A$ and $g_B$ are (efficiently computable) arbitrary functions, and $h$ can be computed in sublinear time on a RAM machine, we get a sublinear communication secure function evaluation protocol for $f$.

**Combining constructions** All our constructions (as well as the garbled circuit transformation) follow the same convention for representing the inputs and outputs: these values are shared between the parties according to a simple secret sharing scheme[5]. A consequence is that in many cases all these constructions may be combined in a single protocol. Hence, we get the following protocol engineering paradigm:

1. Divide a computation into sub-tasks.
2. For each sub-task choose the right computational model (for its insecure computation) that leads to an efficient representation.
3. Transform each sub-task to get its secure function evaluation protocol.
4. Combine these protocols to get a protocol for the computation.

### 1.3 Summary of results

We introduce the *private Indirect Indexing* as well as the *Generalized Indirect Indexing* primitives and show (i) secure evaluation protocols for them (Section 3) (ii) How they can be used to obtain communication preserving protocol for secure evaluation of any function $f(x, y)$ (Section 4) and (iii) How they may be combined with the secure evaluation of circuits (Section 5). We give various examples of problems for which our transformations yields more efficient protocols. In particular we address the millionaires problem – our work provides more efficient protocols for the problem than previously known, both from a communication point of view as well as computation (only a polylogarithmic number of OTs).

Most of the paper deals with the semi-honest model of participants. In Section 6 we discuss a communication efficient method of transforming a protocol in the semi-honest model into one that works in the malicious model.

## 2 Preliminaries

In this work we express complexity in terms of the input length and the cryptographic security parameter[6]. Let $n$ denote the input length and $k$ the security parameter. We consider protocols as *communication efficient* if their communication complexity is no higher than a $\text{polylog}(n) \cdot \text{poly}(k)$ factor times the communication of the best insecure protocol for the same functionality. The security parameter should be selected so that the computational work done by the parties is insufficient for breaking the cryptographic primitives in use. In particular any choice of $k$ must satisfy $k = \omega(\log n)$.

We rely mostly on Goldreich [20, 21] for definitions and notation. In the design and proof of our protocol we make extensive use of *composition theorems* for secure protocols. This has been an active research area (see [11, 20]). An important tool in our design are efficient protocols for Oblivious Transfer (OT) and Private Information Retrieval (PIR). We briefly overview OT and PIR as well as some other cryptographic tools we use.

[5]We use a sharing scheme for bits where a bit is represented by the exclusive or of its shares. Other secret sharing schemes may also be used with minor changes.

[6]This is in contrast to a common practice of identifying the security parameter and the input length.

**Notation** We use $A \stackrel{c}{\equiv} B$ to denote computational indistinguishability of ensembles $A = \{A_n\}_{n\in\mathbb{N}}$ and $B = \{B_n\}_{n\in\mathbb{N}}$ by circuits, so that every family of poly($n$)-size circuits has only a negligible (in $n$) advantage in distinguishing $A$ and $B$. We use neg($n$) to denote functions decreasing faster than any inverse polynomial.

The notation $a \oplus b$ is used for the bit-wise exclusive-or of $a, b$.

## 2.1 The Semi-Honest Model

In most of this work we address the case were the parties participating in the protocol are *semi-honest* i.e. they follow the protocol as prescribed but may record all messages and subsequently deduce information not derivable solely from the protocol output. Let $f : \{0,1\}^* \times \{0,1\}^* \to \{0,1\}^* \times \{0,1\}^*$ be a (randomized) function. Let $\mathcal{P}$ be a two-party protocol for computing $(f_1(x,y), f_2(x,y)) = f(x,y)$. Denote by $\textsc{view}_i^{\mathcal{P}}(x,y)$ the view of the $i$th party during the execution of $\mathcal{P}$ on $(x,y)$ (including her private coins and all received messages) and by $\textsc{out}_i^{\mathcal{P}}(x,y)$ her outcome (which is implicit in $\textsc{view}_i^{\mathcal{P}}(x,y)$).

**Definition 2.1** (Privacy with respect to semi-honest parties) *Protocol* $\mathcal{P}$ privately computes

$$f(x,y) = (f_1(x,y), f_2(x,y))$$

*if there exist probabilistic poly-time simulators* $S_1, S_2$ *such that:*

$$\{S_1(x, f_1(x,y)), f_2(x,y)\} \stackrel{c}{\equiv} \{(\textsc{view}_1^{\mathcal{P}}(x,y), \textsc{out}_2^{\mathcal{P}}(x,y))\}$$
$$\{S_2(x, f_2(x,y)), f_1(x,y)\} \stackrel{c}{\equiv} \{(\textsc{view}_2^{\mathcal{P}}(x,y), \textsc{out}_1^{\mathcal{P}}(x,y))\}$$

**Composition in the semi-honest model** In our constructions we compose protocols that are private with respect to semi-honest parties, and make use of *composition theorems* for secure protocols. In this section we define the notion of private reduction and cite a corresponding composition theorem that enables us to phrase out constructions as composition of protocols. We refer the reader to [20, 11] for further details.

**Definition 2.2** (Privately reducing $g$ to $f$) *An oracle-aided protocol using oracle functionality $f$* privately computes *$g$ if there exist simulators $S_1, S_2$ as in Definition 2.1. The corresponding views are defined in the natural manner to include oracle answers.*

*An oracle-aided protocol* privately reduces *$g$ to $f$ if it privately computes $g$ when using oracle functionality $f$.*

**Theorem 2.1** (Composition in semi-honest model, two parties) *Suppose $g$ is privately reducible to $f$ and there exists a protocol for privately computing $f$. Then, the protocol defined by replacing each oracle-call to $f$ by a protocol that privately computes $f$ is a protocol for privately computing $g$.*

**Forcing semi-honest behavior** The semi-honest model is of a *benign* adversary, that acts as prescribed in the protocol. On the other hand, a *malicious* adversary may deviate arbitrarily from its prescribed protocol. It is known how to transform protocols secure in the semi-honest model into protocols secure in the malicious model [23] (see also [20]). The transformation is via a compiler that 'forces' the parties to act semi-honestly.

A possible design paradigm for secure protocols is thus first to construct a protocol for the semi-honest model and then compile it. Alas, when efficiency is a concern, the compiler is not good enough – in particular it is not communication preserving. In Section 6 we describe an adjustment of the transformation from protocols for the semi-honest model into protocols for the malicious model so as to make it communication preserving.

## 2.2 Cryptographic primitives

We define and discuss the primitives that are used in our constructions. For precise definitions and constructions we refer the reader to [21].

**Pseudo-random generators** A pseudo-random generator is a (deterministic) procedure that transforms a short random seed to a long string that is indistinguishable from a random one. We use pseudo-random generators to generate (pseudo) random bits for a protocol, with the cost of communicating only $k$ bits.

**Definition 2.3** *A pseudo-random generator is a polynomially computable function* $G : \{0,1\}^k \rightarrow \{0,1\}^{poly(n)}$ *so that* $G(U_k) \stackrel{c}{\equiv} U_{poly(n)}$, *where* $U$ *denotes the ensemble of uniform distributions.*

*We denote the computational work for computing* $G(\cdot)$ *by* $W_{PRG}(k)$.

**Pseudo-random functions** A pseudo-random function is a function that is indistinguishable from a random function by an adaptive adversary. A function $F_K : \{0,1\}^n \rightarrow \{0,1\}^{\mathrm{poly}(n)}$ is specified by a short key $K$ of length $k$ bits. For every probabilistic poly-time Turing machine $M$, $|\Pr(M^{F_k}(1^n) = 1) - \Pr(M^H(1^n) = 1)| \leq \mathrm{neg}(n)$, where $H_n$ is a random function mapping $n$ bits to poly($n$) bits. The probability is over the selection of $k$, $H$ and the random coins of $M$.

We denote the computational work for computing $F_k(x)$ by $W_{PRF}(k)$.

### 2.2.1 Oblivious Transfer and Private Information Retrieval

Oblivious Transfer (or OT) is a specific case of secure function evaluation, first suggested by Rabin, where one party (the sender) has some input and the other party (the chooser) learns some aspect of the information without 'hinting' which aspect of the information was transferred.

**Definition 2.4** (1-OUT-OF-$w$ OBLIVIOUS TRANSFER) *Let* $x[0], x[1], \ldots, x[w-1]$ *be elements chosen from* $\{0,1\}^X$. *Let* $j \in \{0, \ldots, w-1\}$ *be an index to one of these elements.*

*An* oblivious transfer *protocol is a protocol that privately computes the function*

$$\mathrm{OT}_1^w((x[0], x[1], \ldots, x[w-1]), j) = (\perp, x[j])$$

*The party holding the index $j$ is referred to as the* chooser *and the party holding the elements $x[0], x[1], \ldots, x[w-1]$ is the* sender.

Oblivious transfer serves as a basic block in our constructions. There are several known constructions of oblivious transfer protocols that are efficient either in communication or in computation (see discussion below). To abstract out the dependency on the specific oblivious transfer protocol in use we express the efficiency of our protocols in terms of its *work* and *communication* denoted $W_{\mathrm{OT}}(w, k, X)$ and $C_{\mathrm{OT}}(w, k, X)$ respectively, where $k$ is the security parameter. We usually ignore the dependency of $W_{\mathrm{OT}}$ and $C_{\mathrm{OT}}$ on $X$ since $X = O(\max(k, \log w))$. Note that since $W_{\mathrm{OT}}(w, k, X) = \Omega(w)$ it follows that the security parameter should be chosen so that $k = \omega(\log w)$, i.e. that breaking the oblivious transfer should take much more work than $w$.

**Complexity of** OT **and PIR** A lot of work was devoted recently to the communication complexity of OT, under the heading of Private Information Retrieval – PIR [6]. The results on *single* PIR by Kushilevitz and Ostrovsky [35] and Cachin et al. [12] solve "half the problem" by protecting the chooser from the sender (i.e. at the end of the protocol the sender cannot distinguish which value the chooser has learned), but the chooser may learn more than a single value. The construction by Kushilevitz and Ostrovsky is under the quadratic residuosity assumption. This was the first sublinear communication protocol for PIR with a *single* database. The scheme by Cachin et al. is under the $\phi$-hiding assumption and is more efficient in terms of communication. The communication complexity of their construction is $k \cdot \mathrm{polylog}(n)$. [7]

In order to protect the sender as well, Naor and Pinkas [41] proposed a method that turns any computational PIR into an $\mathrm{OT}_1^w$ protocol, by applying $\log w$ times (concurrently) an $\mathrm{OT}_1^2$ protocol, without increasing the communication complexity otherwise. Furthermore, they described recently an $\mathrm{OT}_1^w$ protocol based on the Decisional Diffie-Hellman (DDH) assumption [42]. In their protocol the chooser performs a constant number of exponentiations and the sender performs $O(w)$ exponentiations, while the communication complexity is increased by a single element in addition to a PIR scheme[8]. (They also provided work/communication tradeoffs for the problem, which may be relevant in some cases.)

We summarize that under the appropriate assumptions[9] there is an 1-out-of-$w$ oblivious transfer protocol whose communication complexity is proportional to the security parameter plus the size of an element. If one is interested in a low computational protocol, then the overhead can be as little as one exponentiation plus $w$ private-key operations, using the first scheme in [42].

[7] Recently, Kiayias and Yung [37] presented a new polylogarithmic communication PIR scheme.

[8] A similar construction was suggested by Aiello, Ishai and Reingold [1].

[9] $\Phi$-hiding for the PIR scheme [12], and DDH for [42].

**Counting oblivious transfers** The cost of many secure function evaluation protocols is dominated by the cost of the oblivious transfer invocations. Intuitively, the number of oblivious transfer invocations is a good measure for the efficiency of a protocol.

Beaver [4] showed that it is possible to implement poly($k$) oblivious transfers from an initial "seed" of just $k$ oblivious transfers, assuming only the existence of one-way functions (where k is a security parameter). This construction is not efficient since it relies on the secure function evaluation of a pseudo-random generator.

Beaver's result implies that in the *computational setting* (i) in general, specifying the number of oblivious transfer invocation in a protocol is not a sufficient measure for its complexity, and (ii) one cannot hope to obtain a high lower bound (say above logarithmic) on the number of oblivious transfers invocations needed to securely evaluate a functionality.

Nevertheless, in the protocols discussed in this work, the number of oblivious transfer invocations is a dominating factor in complexity.

### 2.3 Garbled circuit secure function evaluation

Many of the works in the field of secure function evaluation followed the 'garbled circuit' construction introduced by Yao [49, 50] and [23]. The idea of the construction is to emulate the (Boolean) circuit for computing $f$ gate by gate (once all the 'garbled' inputs to a gate are known its 'garbled' output may be computed).

The cost of the protocol (e.g. as described in [20]) is of invoking one OT for each wire of the circuit. In a more efficient variant, described e.g. in [43], the cost is that of invoking one $\mathrm{OT}_1^2$ for each input bit plus a constant number of pseudo-random function evaluations per each gate. I.e. the computational work is $n \cdot W_{OT}(2,k) + s \cdot W_{PRF}(k)$ and the communication is $n \cdot C_{OT}(2,k) + O(sk)$. In Appendix A we apply the garbled circuit construction to a *protocol* for computing $f$ (this protocol may leak information but the transformed protocol should not). We give a variant of the garbled circuit transformation that results with computational work $c \cdot W_{OT}(2,k) + s \cdot W_{PRF}(k)$ and communication $c \cdot C_{OT}(2,k) + O(sk)$ where $c$ is the communication complexity of the related (insecure) protocol.

## 3 Indirect Indexing

In the following we describe a primitive called *private indirect indexing* that will serve as a basic building block in all our constructions. The indirect indexing primitive is similar to the oblivious transfer primitive in the sense that in both primitives an indexed array entry is retrieved. The difference is that in the oblivious transfer primitive one party knows the input index $j$ and learns the indexed element $y[j]$ (and the other party learns nothing) whereas in the indirect indexing primitive the input index $j$ and the output $y[j]$ are *shared* between the parties so that (i) the input index is a combination of the parties input shares, and (ii) the protocol results in the parties holding random shares of the indexed element.

**Notation** We use a simple sharing scheme for the inputs and outputs of the indirect indexing primitive. If Alice and Bob hold as shares of $\alpha$ the strings $\pi, \tilde{\pi}$ respectively, then the shared value is $\alpha = \pi \oplus \tilde{\pi}$. It follows that $\tilde{\pi} = \alpha \oplus \pi$ hence we abuse notation and write the shares as $\pi, \alpha \oplus \pi$ (equivalently $\alpha \oplus \tilde{\pi}, \tilde{\pi}$).

Let $\bar{y}$ be an array of $w$ values and $j$ be an index of an element $y[j]$ of $\bar{y}$. The inputs for Alice and Bob are shares $\pi \oplus j, \pi$ of the index $j$. The outputs of Alice and Bob are random shares $\pi', \pi' \oplus y[j]$ of $y[j]$. More formally:

**Definition 3.1** (Private indirect indexing) *Let $\bar{y} = y[0], y[1], \ldots, y[w-1]$ be a list of $w$ values where $y[i] \in \{0,1\}^Y$ and $w$ is an integral power of 2. Let $j \in \{0, \ldots, w-1\}$ be an index to an element of $\bar{y}$.*

*A private indirect indexing protocol (from Alice to Bob) is a protocol $\mathcal{P}_{Ind_{AB}}$ that privately computes the function*

$$Ind_{AB}\left(\pi \oplus j, (\pi, \bar{y})\right) = \left((\pi' \oplus y[j]), \pi'\right)$$

*where $\pi \in \{0,1\}^{\log w}$ and $\pi' \in_R \{0,1\}^Y$.*

*Similarly, the protocol $\mathcal{P}_{Ind_{BA}}$ is defined to privately compute the function*

$$Ind_{BA}\left((\pi, \bar{x}), \pi \oplus j)\right) = (\pi', \pi' \oplus x[j])$$

*where $\pi \in \{0,1\}^{\log w}$ and $\pi' \in_R \{0,1\}^X$.*

**Note 3.1** *The initial string $\pi \in \{0,1\}^{\log w}$ may be chosen arbitrarily. Further, $\pi$ does not have to be chosen explicitly. In particular, $\pi$ may be chosen so that $\pi = 0^{\log w}$ (in such a case Alice knows the input index $j$), or so that $\pi \oplus j = 0^{\log w}$ (in which case Bob knows the input index $j$).*

*On the other hand the string $\pi'$, in the outcome of the protocol, must be selected* uniformly at random *from $\{0,1\}^Y$ i.e. neither Alice nor Bob learn new information about $Y[j]$.*

**Note 3.2** *An equivalent definition is $Ind_{AB}\left(\pi \oplus j, (\pi, \bar{y})\right) = \left(\tilde{\pi}, (\tilde{\pi} \oplus y[j])\right)$ where $\tilde{\pi} \in_R \{0,1\}^Y$.*

We now describe a private indirect indexing protocol. The idea underlying our construction is to use an $\mathrm{OT}_1^w$ protocol where Alice acts as the chooser and Bob the sender. Since Alice does not know which index she should choose, Bob permutes the *locations* of his inputs according to $\pi$. He also encrypts his inputs using a random string $\pi'$. This allows Alice to select the right position, without figuring out the "real" value she has received. In more detail:

**Construction 3.1** (Private indirect indexing protocol (from Alice to Bob)) *Let $\mathcal{P}_{Ind_{AB}}\left(\pi \oplus j, (\pi, y[0], y[1], \ldots, y[w-1])\right)$ be the following protocol:*

**Input:** Alice's input is a permuted index $J = \pi \oplus j$. Bob has as input $\bar{y} = y[0], y[1], \ldots, y[w-1]$ and $\pi \in \{0,1\}^{\log w}$.

**Step 1:** Bob chooses $\pi' \in_R \{0,1\}^Y$ and sets $\mathcal{Y}[\pi \oplus i] = \pi' \oplus y[i]$ for $0 \le i < w$.

**Step 2:** Alice and Bob use an $\mathrm{OT}_1^w$ protocol on inputs $J, (\mathcal{Y}[0], \mathcal{Y}[1], \ldots, \mathcal{Y}[w-1])$ so that Alice learns $J' = \mathcal{Y}[J]$.

**Output:** Alice locally outputs $J'$. Bob locally outputs $\pi'$.

The protocol $\mathcal{P}_{Ind_{BA}}((\pi, x[0], x[1], \ldots, x[w-1]), \pi \oplus j)$ is constructed symmetrically.

**Claim 3.1** *$\mathcal{P}_{Ind_{AB}}$ privately computes the function $Ind_{AB}$ with communication costs $C_{OT}(w, k, Y)$, computation costs $O(W_{OT}(w, k, Y))$ and round complexity of the oblivious transfer protocol.*

**Proof** First we show that the protocol outputs the correct output. Bob locally outputs $\pi'$ for a randomly selected $\pi'$. Alice locally outputs $J' = \mathcal{Y}[J] = \pi' \oplus y[j]$.

As for privacy with respect to semi-honest participants, we show that protocol $\mathcal{P}_{Ind_{AB}}$ privately reduces $Ind_{AB}$ to $\mathrm{OT}_1^w$. By Theorem 2.1 this proves the claim. Let Alice's output be $\pi' \oplus y[j]$ and Bob's output be $\pi'$ as in the protocol. We now construct simulators $S_1, S_2$ in accordance with Definition 2.1. The input to simulator $S_1$ consists of $\pi \oplus j$ and a random number $\tilde{\pi} \in_R \{0,1\}^Y$ (corresponding to $\pi' \oplus y[j]$). The simulator outputs $\tilde{\pi}$. Similarly, the input to simulator $S_2$ consists of $\pi, y[0], y[1], \ldots, y[w-1]$ and a random $\pi'$. The simulator outputs $\pi'$. □

**A simple usage example** We can now sequentially compose $\mathcal{P}_{Ind_{AB}}$ and $\mathcal{P}_{Ind_{BA}}$ to get higher levels of indirect indexing. We begin with an example – the function $Ind$ that performs two levels of indirect indexing.

Let $\bar{x} = x[0], x[1], \ldots, x[w_x - 1]$ and $\bar{y} = y[0], y[1], \ldots, y[w_y - 1]$ where $w_x, w_y$ are integral powers of 2. Define

$$Ind((j, \bar{x}), \bar{y}) = (x[y[j]], \perp)$$

where $j$ is an index to the $y$ list (i.e. $0 \le j < w_y$), each $y[i]$ is an index to the $x$ list (i.e. $0 \le y[i] < w_x$) and $x[i] \in \{0,1\}^X$.

The protocol for $Ind$ first invokes the protocol for computing $Ind_{AB}$ (on inputs $j, \bar{y}$) after which Alice and Bob share $y[j]$ and then invokes the protocol for $Ind_{BA}$ (on inputs $\bar{x}$ and these shares) so that they share $x[y[j]]$. In the last step, Bob sends Alice his share of the output so that Alice may locally output $x[y[j]]$.

**Construction 3.2** *Let $\mathcal{P}_{Ind}((j, \bar{x}), \bar{y})$ be the following protocol:*

**Inputs:** Alice has input $\bar{x} = x[0], x[1], \ldots, x[w_x - 1]$ and an index $j$. Bob's input is a list of indices $\bar{y} = y[0], y[1], \ldots, y[w_y - 1]$.

**Step 1:** Alice and Bob jointly compute $(\tilde{\pi}, \tilde{\pi} \oplus j') = \mathcal{P}_{Ind_{AB}}(j, (0^{\log w_y}, \bar{y}))$.

**Step 2:** Alice and Bob jointly compute $(\tilde{\pi}' \oplus j'', \tilde{\pi}') = \mathcal{P}_{Ind_{BA}}((\tilde{\pi}, \bar{x}), \tilde{\pi} \oplus j')$.

**step 3:** Bob sends $\tilde{\pi}'$ to Alice that locally outputs $j''$.

**Claim 3.2** *$\mathcal{P}_{Ind}$ privately computes the function Ind.*

**Proof** It is easy to see that $j'' = x[y[j]]$, hence the outcome of the protocol is $(x[y[j]], \perp)$ as required.

We show that protocol $\mathcal{P}_{Ind}$ privately reduces $Ind$ to $Ind_{AB}$ and $Ind_{BA}$. By Theorem 2.1 this proves the claim. Construct simulators $S_1, S_2$ in accordance with Definition 2.1 as follows:

The input to simulator $S_1$ contains $\bar{x}$ the index $j$ and the outcome $x[y[j]]$. $S_1$ selects $\tilde{\pi} \in_R \{0, \ldots, w_x\}$ and $\tilde{\pi}' \in_R \{0,1\}^X$ and outputs (i) $\tilde{\pi}$ (this corresponds to the response Alice gets from the $P_{Ind_{AB}}$ oracle in Step 1) and (ii) $\tilde{\pi}' \oplus x[y[j]]$ (this corresponds to the response Alice gets from the $P_{Ind_{BA}}$ oracle in Step 2).

The input to simulator $S_2$ is $\bar{y}$. $S_2$ selects $\tilde{\pi} \in_R \{0, \ldots, w_x\}$ and $\tilde{\pi}' \in_R \{0,1\}^X$ and outputs (i) $\tilde{\pi}$ (this corresponds to the response Bob gets from the $P_{Ind_{AB}}$ oracle in Step 1) and (ii) $\tilde{\pi}'$ (this corresponds to the response Bob gets from the $P_{Ind_{AB}}$ oracle in Step 2). □

**Note 3.3** *Running protocol $\mathcal{P}_{Ind}$ without* **Step 3** *results in both parties holding shares of $x[y[j]]$. To change the protocol so that Bob learns the output,* **Step 3** *should be modified so that Alice sends her share to Bob.*

## 3.1 Generalized Indirect Indexing

We now generalize our usage of $P_{Ind_{AB}}$ and $P_{Ind_{BA}}$ to have $c$ levels of indirect indexing, where the indices alternate between the parties (this is a secure analog to pointer jumping). The inputs to the generalized indirect indexing function $GInd$ are indices ordered in $c$ lists[10]: $\bar{y}_1, \bar{x}_2, \bar{y}_3, \bar{x}_4, \ldots, \bar{y}_{c-1}, \bar{x}_c$. Let $\hat{x} = \bar{x}_2, \bar{x}_4 \ldots, \bar{x}_c$ denote the lists held by Alice and $\hat{y} = \bar{y}_1, \bar{y}_3, \ldots, \bar{y}_{c-1}$ denote the lists held by Bob. For odd $1 \le \ell \le c-1$ the list $\bar{y}_\ell$ contains $w_\ell$ elements $y_\ell[0], y_\ell[1], \ldots, y_\ell[w_\ell - 1]$ where $0 \le y_\ell[i] < w_{\ell+1}$. Similarly, for even $2 \le \ell \le c$ the list $\bar{x}_\ell$ contains $w_\ell$ elements $x_\ell[0], x_\ell[1], \ldots, x_l[w_\ell - 1]$ where $0 \le x_\ell[i] < w_{\ell+1}$ for $\ell \ne c$ (i.e. entries of $\bar{y}_\ell$ may serve as indices to $\bar{x}_{\ell+1}$ and, similarly, entries of $\bar{x}_\ell$ may serve as indices to $\bar{y}_{\ell+1}$), and $x_c[i] \in \{0,1\}^{X_c}$.

The initial index $0 \le j_0 < w_1$ is shared between Alice and bob so that Alice holds $\pi_0 \oplus j_0$ and Bob holds $\pi_0$. The outcome of $GInd$ is $x_c[y_{c-1}[\cdots[x_2[y_1[j_0]]]\cdots]]$ shared by both parties. See Figure 1.

[10]Without loss of generality we assume $c$ to be even.

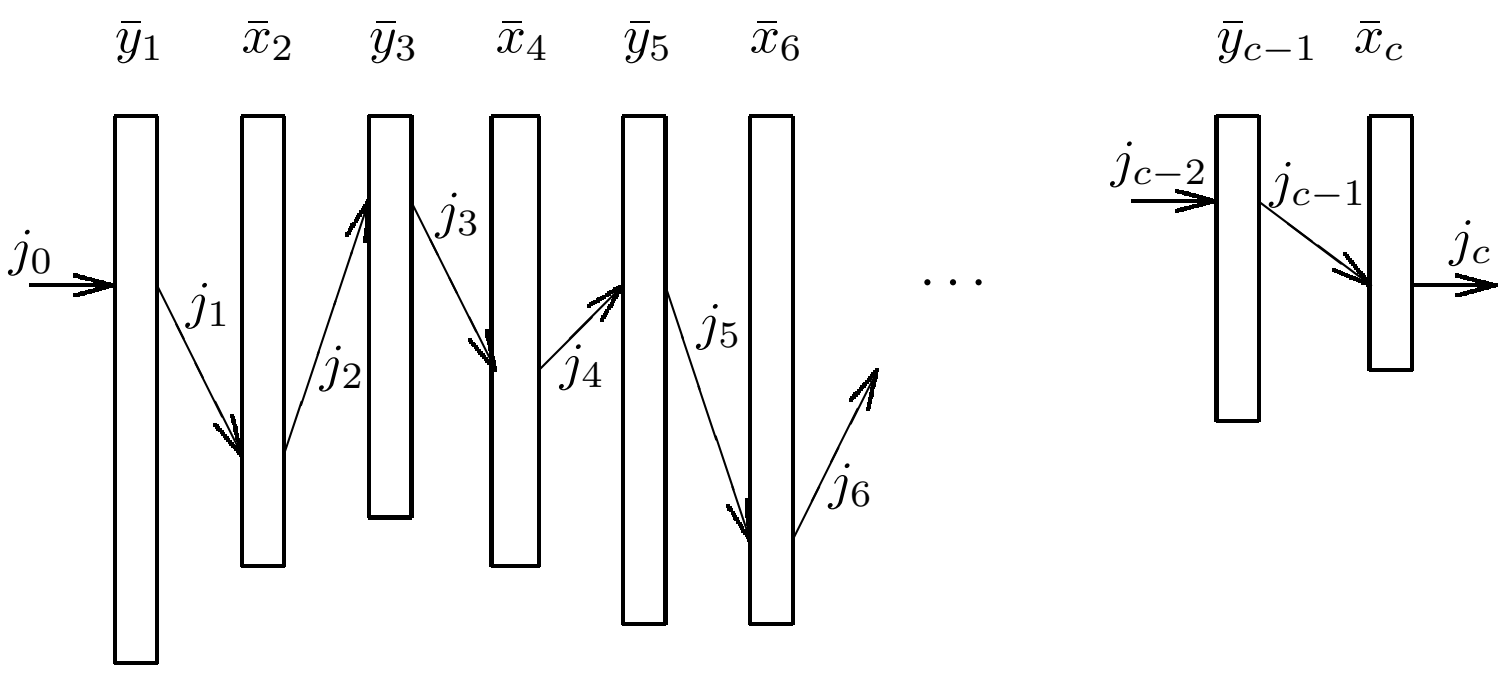

Figure 1: Demonstration of *GInd*. Alice holds $\hat{x} = \bar{x}_2, \bar{x}_4 \ldots, \bar{x}_c$ and Bob holds $\hat{y} = \bar{y}_1, \bar{y}_3, \ldots, \bar{y}_{c-1}$. The outcome is $x_c[y_{c-1}[\cdots[x_2[y_1[j_0]]]\cdots]]$ shared between Alice and Bob. The intermediate indices $j_1, \ldots, j_c$ are not revealed to any of the parties.

**Definition 3.2** (Private general indirect indexing) *For* $1 \leq \ell \leq c$ *let*

$$j_\ell = \begin{cases} y_\ell[j_{\ell-1}] & \text{if } \ell \leq c \text{ is odd} \\ x_\ell[j_{\ell-1}] & \text{if } \ell \leq c \text{ is even} \end{cases}$$

*Then*

$$GInd((\pi \oplus j_0, \hat{x}), (\pi, \hat{y})) = (\pi' \oplus j_\ell, \pi')$$

*where* $\pi' \in_R \{0,1\}^{X_c}$.

*A private general indirect indexing protocol is a protocol that privately computes the function GInd.*

The construction of $\mathcal{P}_{GInd}$ is similar to that of $\mathcal{P}_{Ind}$ (Construction 3.2). It uses $\mathcal{P}_{Ind_{AB}}$ and $\mathcal{P}_{Ind_{BA}}$ alternately. In more detail:

**Construction 3.3** *Let* $\mathcal{P}_{GInd}((\pi \oplus j_0, \hat{x}), (\pi, \hat{y}))$ *be the following protocol:*

**Inputs:** Alice's input is $\hat{x} = \bar{x}_2, \bar{x}_4 \ldots, \bar{x}_c$ and a masked index $\pi \oplus j_0$. Bob's input is $\hat{y} = \bar{y}_1, \bar{y}_3, \ldots, \bar{y}_{c-1}$ and the string $\pi$. Let $\pi_0 = \pi$.

**Step** $1 \leq \ell \leq c-1$ **odd:** Alice and Bob jointly compute

$$(\pi_\ell, \pi_\ell \oplus j_\ell) = \mathcal{P}_{IndAB}(\pi_{\ell-1} \oplus j_{\ell-1}, (\pi_{\ell-1}, \bar{y}_\ell)).$$

**Step** $2 \leq \ell \leq c$ **even:** Alice and Bob jointly compute

$$(\pi_\ell \oplus j_\ell, \pi_\ell) = \mathcal{P}_{Ind_{BA}}((\pi_{\ell-1}, \bar{x}_\ell), \pi_{\ell-1}(j_{\ell-1})).$$

The proof of correctness and privacy is similar to that of claims 3.1 and 3.2. Using the notation of Section 2.2.1 we get:

**Theorem 3.3** *Protocol* $\mathcal{P}_{GInd}$ *privately computes function GInd with work* $O(\sum_{\ell=1}^{c} W_{\mathrm{OT}}(w_\ell, k))$ *and communication* $O(\sum_{\ell=1}^{c} C_{\mathrm{OT}}(w_\ell, k))$.

# 4 Communication Preserving Protocols

As a consequence of Theorem 3.3 we give two transformation from (insecure) protocols into communication preserving secure function evaluation protocols. These transformations formulate our first design methodology for secure function evaluation protocols, and we demonstrate how to apply them to several examples.

**Overview of the section** We begin by considering protocols in the communication complexity model (for a more detailed account of the model see [34]). Applying Theorem 3.3 to such protocols (represented by communication complexity trees) we get our main theorem – a communication preserving transformation into secure function evaluation protocols. A naive application of our transformation to protocols may result in protocols with super-polynomial work. Hence, we give a variant of our main theorem using a possibly more succinct representation, by oblivious branching programs, that allows for less work. Our transformations are formulated with deterministic protocols. We discuss how to use randomized protocols with these transformation, using fairly standard tools from communication complexity and cryptography. Application examples are given throughout the section.

## 4.1 Communication Complexity Protocols

Intuitively, communication complexity protocols are protocols between two parties who wish to evaluate a function of their joint input. They engage in a protocol, where they alternately send single bit messages[11]. The protocol results in the computation of the function by one of the parties (or both). The main issue in the communication complexity model is efficiency, in terms of the amount of communication needed to jointly compute the function. We emphasize that security is not an issue in this model, and will be introduced by our transformations.

Assume two players, Alice and Bob, wish to evaluate $f(x, y)$ where Alice's input is $x$ and Bob's input is $y$. Without loss of generality we assume that Alice sends the first message of the protocol. Note that her input $x$ determines her message. Furthermore, her input determines her next message for any sequence of messages she may get from Bob. Similarly, Bob's input $y$ determines his next message for any message sequence he may get from Alice. Hence, upon seeing their respective inputs Alice and Bob may choose their next message for any possible message sequence. Note that this may be done before the protocol is actually run. We note that in the protocols we deal with, the number of messages sent by Alice and Bob depends only on the input length $n$ (and not on the actual value of $x, y$).

The layout of the protocol is of a full binary tree, where every internal node $v$ is labeled by a function ($a_v(\cdot)$ or $b_v(\cdot)$), and every leaf $v$ is labeled by a value $z_v$. Every node $v$ corresponds to a possible message sequence in the natural manner, according to the path from the root to $v$ (i.e. choosing a left edge corresponds to sending the message "0", choosing a right edge corresponds to sending the message "1"). Following our convention that Alice sends the first message, she controls the tree edges from even to odd depth nodes, and Bob controls the

[11]The case where the parties send longer messages is analogous assuming the protocol is oblivious in the sense that for every round the number of bits that are sent does not depend on the input $(x, y)$.

other edges. Upon seeing her input, Alice decides for each internal node $v$ of even depth whether to choose a right or left edge. This choice is reflected in the definition below by the functions $a_v$ applied to Alice's input $x$. Similarly, Bob decides for each internal node $v$ of odd depth whether to choose a right or a left edge. This is reflected in the definition by the functions $b_v$ applied to Bob's input $y$. Each leaf $v$ is assigned a value $z_v$, and the outcome of the protocol is the value of the leaf that is reached by the alternate choices of Alice and Bob. More formally:

**Definition 4.1** (Communication complexity model) *A protocol $\mathcal{P}_{CC}$, in the communication complexity model, over domain $X \times Y$ with range $Z$ is a full binary tree of even height where (i) Each internal node $v$ of even depth is labeled with a function $a_v : X \to \{0,1\}$, (ii) Each internal node $v$ of odd depth is labeled with function $b_v : Y \to \{0,1\}$, and (iii) Each leaf $v$ is labeled with an element $z_v \in Z$.*

*The* value *of the protocol $\mathcal{P}_{CC}$ on input $(x, y)$ is the label $z_v$ of the leaf reached by starting at the root, and walking on the tree, according to the functions $a_v, b_v$. The protocol computes $f : X \times Y \to Z$ if its value on input $(x, y)$ equals $f(x, y)$. The* cost *$c$ of $\mathcal{P}_{CC}$ is the height of the tree.*

Note that the labels $a_v(\cdot), b_v(\cdot)$ and $z_v$ are known a-priori to both parties. The outcome of $a_v(x)$ and $b_v(y)$ is computable by Alice and Bob respectively upon seeing their input. To actually run the protocol, Alice constructs (perhaps implicitly) from $\mathcal{P}_{CC}$ the labeled sub-tree induced by her input, by hard-wiring the choices $a_v(x)$ for edges from even to odd depth nodes. Similarly, Bob hard-wires the choices $b_v(y)$ for edges from odd to even depth nodes.

### 4.1.1 Example: A protocol for the Hamming distance

We consider a concrete example – a protocol for computing the Hamming distance of two $n$-bit strings $x, y$ (i.e. the number of locations $i$ such that $x_i \neq y_i$). The protocol proceeds as follows. Alice sends her input $x$ to Bob bit by bit starting with $x_0$. For each bit $x_i$ of $x$ Bob receives he checks whether $y_i$ agrees with $x_i$. Bob replies with a 0-message if $y_i = x_i$ and with a 1-message if $y_i \neq x_i$. At the end of the protocol, Alice and Bob may compute the Hamming distance of $x, y$ by counting the number of 1-messages sent by Bob. The cost of the protocol is of $c = 2n$ single bit messages.

**The protocol tree** We give a more formal description of the protocol tree. We label every node $v$ of the tree in the natural manner – with a binary number corresponding to the sequence of left (0) and right (1) edges on the path from the root to $v$ (intuitively, this corresponds to a sequence of single bit messages $\sigma_0, \sigma_1, \ldots$ leading to $v$). The root is labeled $\epsilon$. We now define the functions $a_v(\cdot), b_v(\cdot)$ and the values $z_v$. The functions $a_v$ determine Alice's messages, hence, for every internal node of even depth $v = \sigma_0, \ldots, \sigma_{2i-1}$ we set

$$a_v(x) = x_i.$$

This setting of $a_v$ corresponds to Alice sending the bit $x_i$. Similarly, the functions $b_v$ determine Bob's messages. Here the situation is slightly more complicated since Bob's messages depend on the last message sent by Alice and on his input. Hence, for every internal node of odd depth $v = \sigma_0, \ldots, \sigma_{2i}$ we set

$$b_v(y) = \begin{cases} 0 & \text{if } y_i = \sigma_{2i} \\ 1 & \text{otherwise} \end{cases}$$

This setting of $b_v$ corresponds to Bob sending a 0-message if $y_i = x_i$ and a 1-message otherwise. Finally we define the leaf values $z_v$ to count the number of 1-messages sent by Bob and let

$$z_v = \sum_{i=0}^{n-1} \sigma_{2i+1}$$

for every leaf $v = \sigma_0, \ldots, \sigma_{2n-1}$. Figure 2 (i) displays the protocol layout for two-bit inputs. The nodes are numbered from left to right, corresponding to the labels above.

**Note 4.1** *We emphasize that the protocol tree, as described above, is 'fixed' in the sense that it does not depend on the actual inputs of the parties. In particular, we may assume that the protocol tree is agreed upon and known to both parties prior to running the protocol.*

**Running the protocol** We separate the running of the protocol into two stages[12] *preprocessing* and *communication*, as follows:

In the preprocessing stage, Alice and Bob make their decisions for the entire protocol by computing $a_v(x)$ and $b_v(y)$ respectively. We emphasize that no communication is needed for this computation, since each party computes functions of its input alone. The outcome of the computation is that each party holds an induced subtree of the protocol tree, where Alice removes from her tree the edges not chosen by the functions $a_v(x)$, and Bob removes those not chosen by $b_v(y)$. Figures 2 (ii) and (iii) show Alice's subtree for the input $x = 01$ and Bob's subtree for the input $y = 11$.

**Note 4.2** *In the following we do not give an explicit account of the computational work needed for computing $a_v(x)$ and $b_v(y)$. It can be checked, however, that this work is feasible for all our examples.*

In the communication stage, Alice and Bob follow a path from the root to a leaf in their subtrees. The first edge is determined by Alice's subtree – it is the (only) edge exiting the root in her subtree. Alice sends a 0-message if this edge is a left edge and a 1-message otherwise. The path continues by alternately following the edges in Bob's and Alice's subtree, until a leaf is reached. The outcome of the protocol is the leaf value.

As an example, consider the actual run of the Hamming distance protocol on inputs 01 and 11. Alice's first message is $a_\varepsilon(01) = 0$, this corresponds to sending her first bit. Bob's reply is $b_0(11) = 1$, i.e. a disagreement. Alice then sends $a_{10}(01) = 1$ (her second bit) and Bob replies $b_{110}(11) = 0$ (agreement) and the outcome is $z_{0110} = 1$. Note how this actual run is reflected in the subtrees held by Alice and Bob: Alice's first message corresponds to the (left) edge exiting the root in (i); Both Alice and Bob know now that node 0 of depth 1 was reached. Bob's reply corresponds to the (right) edge exiting this node in (ii), reaching node 1 of depth 2. Alice's second message then corresponds to the (right) edge exiting this node in (i), reaching node 3 of depth 3. Finally Bob's last message corresponds to the (left) edge exiting this node in (ii), reaching a leaf labeled 1 (the protocol outcome).

[12] This separation into two stages is not required in an actual run of the protocol.

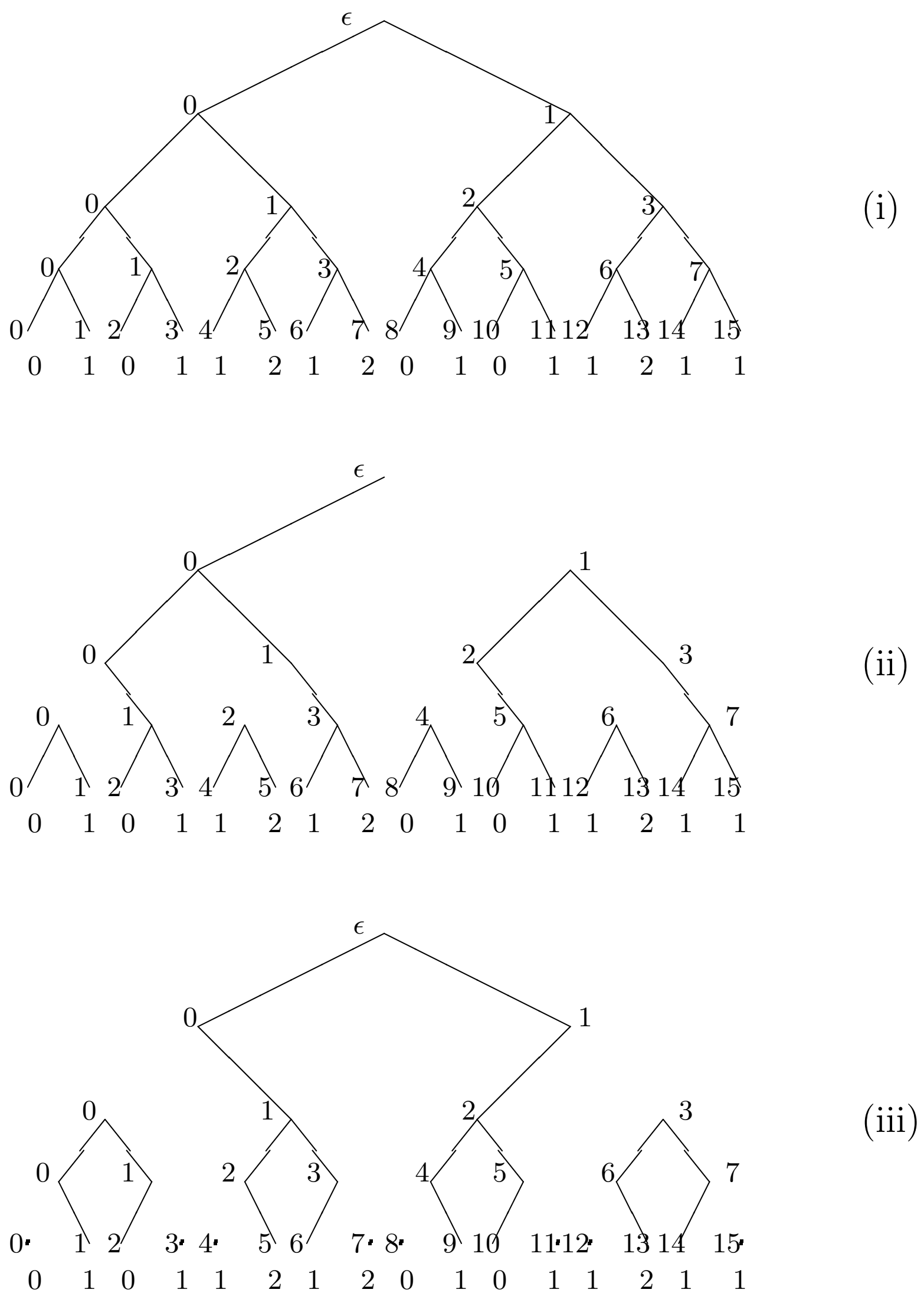

Figure 2: Protocol for computing the Hamming distance of $x, y$.

## 4.2 Secure function evaluation based on the communication complexity tree

We now turn to translate the problem of evaluating a protocol $\mathcal{P}_{CC}$ into an instance of $GInd$, hence resulting in a secure evaluation protocol for $\mathcal{P}_{CC}$. Let $\mathcal{P}_{CC}$ be a protocol for computing $f$ in the communication complexity model, with cost $c$. For every node $v$ in level $\ell$ let $n(v)$ be its position from the left (i.e. for the leftmost node on the $\ell$th level $n(v) = 0$ and for the rightmost $n(v) = 2^\ell - 1$). We now reduce the problem of computing $\mathcal{P}_{CC}$ to the problem of general indirect indexing. The idea of the reduction is that one can view any odd step $\ell$ of a protocol in the communication complexity model as if Alice chooses from the $2^\ell$ labels induced by Bob's input, followed by Bob choosing from the $2^{\ell+1}$ labels induced by Alice's input. The choices are made based on the transcript of the protocol so far and the bits Alice and Bob send in step $\ell$. The protocol $\mathcal{P}_{CC}$ reduces to an instance of *GInd* with $c$ levels, where the $\ell$th level has an array of size $w_\ell = 2^\ell$, whose values are the labels of the corresponding nodes in the tree.

The parties represent their subtrees by a sequence of lists $\bar{y}_1, \bar{x}_2, \bar{y}_3, \bar{x}_4 \ldots, \bar{y}_{c-1}, \bar{x}_c$, and an initial index $j$.

- Bob's subtree is represented by the lists $\bar{y}_1, \bar{y}_3, \ldots, \bar{y}_{c-1}$ as follows:
  - For a node $v$ being the $i$th node of (odd) depth $\ell$, let $y_\ell[i]$ be the node (of depth $l+1$) that is reached by taking the edge exiting from $v$ in Bob's subgraph.
- Alice's subtree is represented by $j, \bar{x}_2, \bar{x}_4, \ldots, \bar{x}_{c-2}$ as follows:
  - The initial index $j$ equals the node (of depth 1) reached by the taking the edge exiting the root in Alice's subtree.
  - For a node $v$ being the $i$th node of (even) depth $\ell$, let $x_\ell[i]$ be the node (of depth $l+1$) that is reached by taking the edge exiting from $v$ in Alice's subgraph.
- The leaf values are represented by the list $\bar{x}_c$ as follows:
  - For a node $v$ being the $i$th leaf let $x_c[i] = z_v$.

It follows that going by the lists, starting with the initial index $j$ we get the outcome of the protocol

$$x_c[y_{c-1}[\cdots[x_2[y_1[j]]\cdots]].$$

**Construction 4.1** (Private computation of $\mathcal{P}_{CC}$) *Let* $\mathcal{P}_{CC}^{\mathrm{PRIV}}(x, y)$ *be the following protocol:*

**Inputs:** Alice holds input $x \in X$. Bob holds input $y \in Y$.

**Step 1:** Let $w_\ell = 2^\ell$.

Alice sets $j = a_{root}(x)$. For all nodes $v$ in even level $2 \le \ell < c$ she sets $x_\ell[n(v)] = 2n(v) + a_v(x)$. For all leaves $v$ she sets $x_c[n(v)] = z(v)$.

For all nodes $v$ in odd level $1 \le \ell \le c-1$ Bob sets $y_\ell[n(v)] = 2n(v) + b_v(y)$.

**Step 2:** Alice and bob run protocol $\mathcal{P}_{GInd}$ on $((j, \hat{x}), \hat{y})$ where $\hat{x} = \bar{x}_2, \bar{x}_4, \ldots, \bar{x}_c$ and $\hat{y} = \bar{y}_1, \bar{y}_3, \ldots, \bar{y}_{c-1}$.

We conclude the following theorem:

**Theorem 4.1** *Let $\mathcal{P}_{CC}$ be a protocol in the communication complexity model computing $f : X \times Y \to Z$ with cost c. Then there exists a protocol $\mathcal{P}_{CC}^{\mathrm{PRIV}}$ privately computing $f$ with $O(c)$ rounds, $O(c \cdot C_{\mathrm{OT}}(2^c, k))$ communication and $O(W_{\mathrm{OT}}(2^c, k))$ work.*

**Note 4.3** *Theorem 4.1 holds also for the case where the parties are not limited to sending a single bit at every round, as long as the number of bits sent at each round does not depend on the inputs of the parties. Therefore one can also have a round preserving protocol.*

**Example** Applying the Construction 4.1 to the Hamming distance protocol in Figure 2 we get:

$$\begin{aligned} &j = 0 \\ &\bar{y}_1 = [1, 2] \\ &\bar{x}_2 = [1, 3, 5, 7] \\ &\bar{y}_3 = [1, 2, 5, 6, 9, 10, 13, 14] \\ &\bar{x}_4 = [0, 1, 0, 1, 1, 2, 1, 2, 0, 1, 0, 1, 1, 2, 1, 2]. \end{aligned}$$

Going through the lists we get:

$$x_4[y_3[x_2[y_1[j]]]] = x_4[y_3[x_2[y_1[0]]]] = x_4[y_3[x_2[1]]] = x_4[y_3[3]] = x_4[6] = 1.$$

### 4.2.1 Simple applications

**Computing the median** Assume Alice and Bob hold subsets $x, y$ of $\{1, \ldots, n\}$. They would like to compute the median of the multi-set $x \cup y$. There exists a protocol for the median problem with communication complexity $O(\log n)$ (due to M. Karchmer, see [34] Ex. 1.7). Hence Theorem 4.1 yields a communication preserving protocol for the median, with polynomial work.

We give a short description of the protocol. For simplicity, we assume that $x$ and $y$ have the same size $m$, which is a power of 2. The protocol runs in $O(\log n)$ steps. Alice and Bob maintain subsets of $x, y$ so that their size is cut by half at every step as follows. Let $x_i, y_i$ be the subsets of $x, y$ in the $i$th step. Denote the medians of $x_i, y_i$ by $m_a, m_b$. To reduce the size of their subsets, Alice and Bob compare $m_a, m_b$. If $m_a = m_b$ then the median is found. Otherwise, Alice and Bob reduce the sizes of their sets by half. If $m_a < m_b$ Alice removes the smallest elements of her set and Bob removes the greatest elements of his set. Similarly, if $m_a > m_b$, Alice removes the greatest elements of her set and Bob removes the smallest elements of his set.

The crux of the protocol is an efficient comparison of $m_a, m_b$ so that the total communication cost is $O(\log n)$. The medians $m_a, m_b$ are compared single bit at a time starting from the high order bits, stopping on a bit $j$ where $m_a[j], m_b[j]$ disagree. We will show that the high order bits should not be exchanged in the following steps of the protocols, resulting with a protocol with logarithmic communication complexity. Consider for example the case $m_a < m_b$. After removing the half smallest element in her set, all the remaining elements in Alice's set have the same high order bits as $m_a$, hence Alice does not have to re-send them in the following steps. A similar argument holds for the remaining elements in Bob's subset.

**Comparing (deterministic) strategies** Let $G$ be some combinatorial game that takes $d$ steps and assume that Alice and Bob each have a (deterministic) strategy for $G$, explicitly described as a binary game tree of depth $d$ and total size $n = 2^d$. They wish to determine whose strategy is winning, without revealing to the other party any other information. To decide whose strategy is winning they may play the game for $d$ steps (and $d$ bits of communication). However, this reveals more than they want. A protocol based on a circuit for comparing the strategies results in costs of invoking an $\mathrm{OT}_1^2$ protocol $\Omega(n)$ times. On the other hand, a protocol in the communication complexity model for comparing the strategies has cost proportional to the depth $d$ of the tree and hence the communication costs if such a protocol is $O(d \cdot C_{\mathrm{OT}}(n,k))$ and the work is $O(d \cdot W_{\mathrm{OT}}(n,k))$.

## 4.3 Oblivious Branching Program Protocols

One problem with the explicit description of a function by its communication complexity tree is that the amount of work done by the two parties is proportional to $2^c$, which can be large, once the communication $c$ is super-logarithmic. We therefore consider a representation of a protocol that may allow for less work, by restricting the amount of information needed to be stored to execute the protocol. The communication tree of Definition 4.1 gives for each participant a tree, induced by its input. Suppose that instead of a tree each participant has a DAG, or an oblivious branching program, i.e. all *equivalent* nodes of a given level (in the sense that there is no input of the other party that is consistent with the current state and that will make the two nodes decide a different value for the function) are merged into one node. The important parameter is the *width* $w$, the maximum number of nodes at a level, of the resulting program[13].

It is possible to define the branching program model as if each party holds a branching program (represented by a labeled layered directed graph) determined by his input. Instead, we give a definition which is a bit more convenient to convert into a secure protocol. In our definition the parties maintain a single branching program, where Alice controls the edges from even depth to odd depth, and Bob controls the edges from odd depth to even depth (this is similar to our definition of the communication complexity tree). [14] Note that any protocol $\mathcal{P}_{CC}$ of communication complexity $c$ and width $2^c$ as defined above can be put in the form of Definition 4.2 below with similar width and cost.

**Definition 4.2** (Oblivious branching program model) *A protocol $\mathcal{P}_{BP}$, in the (oblivious) branching program model, over domain $X \times Y$ with range $Z$ is a layered directed graph $G = (V = (L_0, L_1, L_2, \ldots, L_c), E)$ of even depth $c$ where (i) $|L_0| = 1$, (ii) Each internal node $v$ in layer of even depth $\ell < c$ is labeled with a function $a_v : X \to L_{\ell+1}$, (iii) Each internal node $v$ in layer of odd depth $\ell < c$ is labeled with a function $b_v : Y \to L_{\ell+1}$, and (iii) Each node in $L_c$ is labeled with an element $z \in Z$.*

[13] Beame et al [10] proposed a different notion of communicating branching programs - they charged the participants for access to their own input as well, whereas in our case this is "for free."

[14] The resulting width in our definition may be in the worst case the square of the width in the model where each party holds a branching program. However, parties often "share" states, hence the width is much lower.

*The* value *of the protocol $\mathcal{P}_{BP}$ on input $(x, y)$ is the leaf reached by starting at the single element of $L_0$ and walking on the graph according to the functions $a_v, b_v$. The protocol computes $f : X \times Y \to Z$ if its value on input $(x, y)$ equals $f(x, y)$. The* cost *of $\mathcal{P}_{BP}$ is $c$, its* width *is* $\max |L_\ell|$.

We now use use our generalized indirect indexing protocol to privately compute $\mathcal{P}_{BP}$. The construction is similar to that of $P_{CC}^{\mathrm{PRIV}}$. In the following the function $n(\cdot)$ enumerates states so that $\{n(v)|v \in L_\ell\} = \{0, \ldots, |L_\ell| - 1\}$.

**Construction 4.2** (Private computation of $\mathcal{P}_{BP}$) *Let $\mathcal{P}_{BP}^{\mathrm{PRIV}}(x, y)$ be the following protocol:*

**Inputs:** Alice holds input $x \in X$. Bob holds input $y \in Y$.

**Step 1:** Let $w_\ell = |L_\ell|$.

Alice sets $j = a_{root}(x)$ where *root* is the single element in $L_0$. For all nodes $v$ in even level $\ell < c$ she sets $x_\ell[n(v)] = n[a_v(x)]$. For all leaves $v$ she sets $x_c[n(v)] = z(v)$.

For all nodes $v$ in odd level $\ell$ Bob sets $y_\ell[n(v)] = n[b_v(y)]$.

**Step 2:** Alice and bob run protocol $\mathcal{P}_{GInd}$.

**Theorem 4.2** *Let $\mathcal{P}_{BP}$ be a protocol in the branching program model computing $f : X \times Y \to Z$ with cost $c$ and width $w$. Then there exists a protocol $\mathcal{P}_{BP}^{PRIV}$ privately computing $f$ with $O(c)$ rounds, $O(c \cdot C_{\mathrm{OT}}(w, k))$ communication and $O(c \cdot W_{\mathrm{OT}}(w, k))$ work.*

**Comparing Strings** We give a fairly detailed construction of a string comparison branching program protocol. On input two $w$-bit strings $x, y$, let $f = 1$ if $x = y$ and $f = 0$ otherwise. Figure 3 (i) depicts a branching program computing $f$ corresponding to the protocol where Alice and Bob exchange their strings, one bit at a time. The branching program checks whether Alice and Bob disagree some bit. Once a disagreement is found (i.e. $x_\ell \neq y_\ell$), the program moves into a 'trap' state that leads, eventually, to the leaf labeled 0. Otherwise the leaf labeled 1 is reached. For inputs $x = 1101$ and $y = 1010$, Figure 3 (ii) and (iii) shows the corresponding branching programs held by Alice and Bob.

**Note 4.4** *In a typical usage, the strings $x, y$ are hashed prior to applying the branching program of Figure 3, where the hash function is selected in the clear (see discussion in Section 4.4.1 and usage in Section 4.5.1).*

**Acceptance by an automaton** Alice holds an deterministic finite automaton $\mathcal{A} = (Q, \delta)$ where $Q = \{q_0, q_1, \ldots, q_n\}$ and $\delta : Q \times \{0, 1\} \to Q$. Bob holds a string $\alpha \in \{0, 1\}^c$. They wish to decide whether $\alpha$ is accepted by $\mathcal{A}$ (i.e. whether $\alpha$ reaches $q_n$ starting from $q_0$).

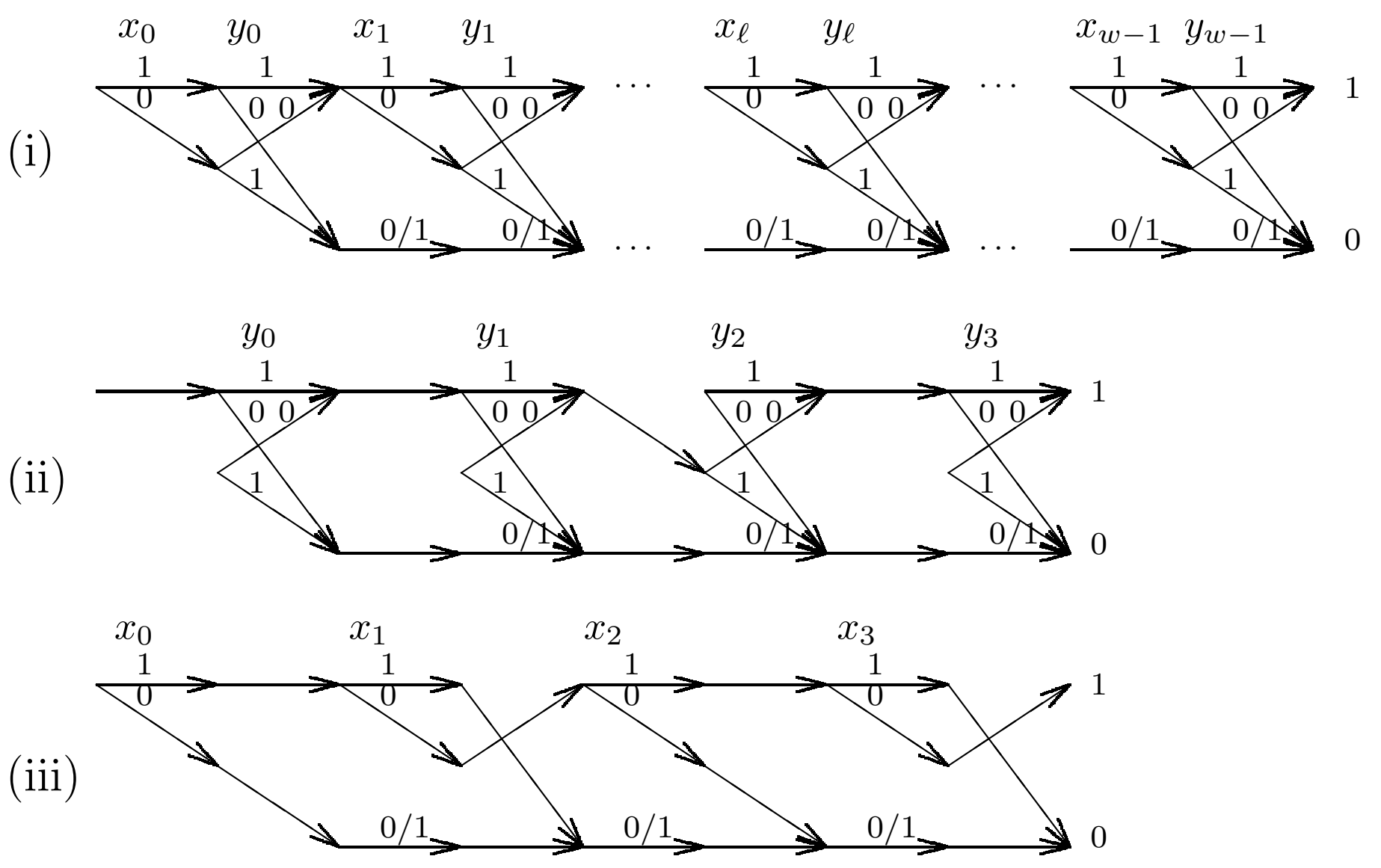

Figure 3: Branching program protocol for string comparison

Define a branching program based on $\mathcal{A}$ with cost $O(c)$ and width $O(n)$ (this branching program is degenerate in the sense that it is completely defined by Alice). The program considers a single bit of $\alpha$ at a time and stores the state reached by the prefix of $\alpha$. Let $L_0 = \{q_0\}$ and $L_i = Q$ for $0 < i \leq c$. Label node $q_n$ in $L_c$ as 1 and all other nodes as 0. For each node $q \in L_i$, the function $a_q$ is defined to be $a_q(\cdot) = \delta(q, \cdot)$. It is easy to see that the branching program decides whether $\mathcal{A}$ accepts $\alpha$.

## 4.4 The Randomized Case

The use of randomized protocols allows, in many interesting cases, to reduce the communication complexity of protocols to be below linear in the input size. We present two approaches that allow the use of such protocols with theorems 4.1 and 4.2. The approaches differ in their treatment of random coins. In one, the random bits are revealed to both parties, and in the other they are kept secret from the parties.

In the *revealed random bits* approach (Lemma 4.3) we start with a protocol with a very small probability of error and 'derandomize' it by selecting random coins (independent of the input) and 'hard-wiring' them into the protocol. Theorem 4.1 (or Theorem 4.2) is then applied to the 'derandomized' protocol. In *secret random bits* approach we start with a protocol with low randomness.[15] The protocol is changed to a new protocol where each party initially holds a share of the random string (according to some secret sharing scheme). The protocol first retrieves the random string and then executes the original protocol with respect to this string. Theorem 4.1 or Theorem 4.2 may be applied to this protocol to obtain a private one. A drawback of this approach is that in general it results with super polynomial work protocols.

[15]Lemma 4.4 serves as a standard tool to reduce the randomness of every protocol to be roughly logarithmic in $n$ and in the error probability.

Our examples in Section 4.5 exemplify a design paradigm that combines both approaches. We compute a function in 'parts', so that (i) each part has an efficient private implementation (ii) at the end of each part the parties share only a small amount of secret random bits (needed e.g. to conceal intermediate results), and (iii) the overall computation, given the outputs of all parts is private and efficient.

We begin by a brief introduction of randomized protocols. We refer the reader to [34] for a detailed account of randomized protocols.

A *private-coin* randomized protocol is a protocol as in Definitions 4.1 or Definition 4.2 in which Alice has access to a random string $r_A$ and Bob has access to a random string $r_B$. The two strings are chosen independently uniformly at random. In the tree (or branching program) defining the protocol Alice's nodes are labeled by functions $a_v$ of $x$ and $r_A$; Bob's nodes are labeled by functions $b_v$ of $y$ and $r_B$. A *public-coin* randomized protocol is defined similarly where both Alice and Bob has access to the same random string $r$. We now define what it means for a protocol to compute a function.

**Definition 4.3** *Let $\mathcal{P}, \mathcal{P}'$ be private-coin and public-coin randomized protocols respectively. $\mathcal{P}$ computes a function $f$ with error $\varepsilon$ if $\Pr_{r_A,r_B}[\mathcal{P}((x, r_A), (y, r_B)) = f(x, y)] \geq 1-\varepsilon$ for every $(x, y)$. Similarly, $\mathcal{P}'$ computes a function $f$ with error $\varepsilon$ if $\Pr_r[\mathcal{P}'((x, r), (y, r)) = f(x, y)] \geq 1 - \varepsilon$ for every $(x, y)$.*

### 4.4.1 Revealed random bits transformation

We now turn to describe our *revealed random bits* transformation from randomized protocols to private protocols. We first observe that if the probability of error is negligible, then the randomized case is not significantly different than the deterministic case. As long as Alice and Bob choose their random coins independently of the input, the chances of choosing 'bad' coins so that the protocol outcome differs from $f$ are negligible. Note that even if Alice and Bob know the random string with which the protocol runs, they do not gain any significant information (in a computational sense) from it. It follows that Alice and Bob may 'derandomize' the protocol by choosing a random string and 'hard-wiring' it into the protocol. To save on communication they select a short random seed and expand it using a pseudo-random generator. The work and communication of the resulting protocol are similar to those obtained by theorems 4.1 and 4.2.

**Lemma 4.3** *Let $\mathcal{P}$ be a randomized protocol computing (deterministic) function $f : X \times Y \rightarrow Z$ with cost $c$ and negligible error probability. Let $G$ be a pseudo-random generator as in Definition 2.3. Then there exists a protocol $\mathcal{P}'$ privately computing $f$ with costs as in theorems 4.1 and 4.2 plus the cost of communicating $k$ random bits.*

**Proof** The construction of $\mathcal{P}'$ is based on the observation that the outcome of $\mathcal{P}$ is computationally close to the outcome of an ideal deterministic procedure that correctly computes $f$. It follows that if both Alice and Bob 'hard-wire' pseudo-random coins into the protocol, then with extremely high probability none of them will learn from the outcome of the induced protocol more than it is possible to learn from the outcome of the ideal procedure. It is possible to apply Theorem 4.1 or Theorem 4.2 on the induced protocol to get a private protocol.

We privately reduce $f$ to $\mathcal{P}_{CC}^{\mathrm{PRIV}}$ (or to $\mathcal{P}_{BP}^{\mathrm{PRIV}}$). By Theorem 2.1 this proves the lemma. Let $G$ be a pseudo-random generator. Consider the following protocol:

**Construction 4.3**

**Step 1:** Alice selects a seed $s \in_R \{0,1\}^k$ for $G$ and sends $s$ to Bob. The string $r = G(s)$ defines the private and public coins and induce a deterministic protocol $\mathcal{P}^r$ where $a_v^r(x) = a_v(x, r)$ and $b_v^r(y) = b_v(y, r)$.

**step 2:** Alice and Bob apply Theorem 4.1 (or Theorem 4.2) on $\mathcal{P}^r$.

Construct a simulator $S_1$ for Alice in accordance with Definition 2.1. Note that with extremely high probability (over the choice of $r$), the outcome of the protocol is $(\pi \oplus f(x,y), \pi)$ as required. Thus we can assume the input of $S_1$ to be $x, \pi \oplus f(x,y)$. The simulator $S_1$ chooses $s$ at random and outputs $G(s), \pi \oplus f(x,y)$. The simulator $S_2$ is constructed similarly. □

### 4.4.2 Secret random bits transformation

The *secret random bits* transformation applies to the more general case, e.g. when the computed functionality is a distribution. The drawback of this approach is that it is not guaranteed to yield protocols with polynomial work.

We begin with a standard lemma that states that any randomized protocol can be transformed into a public-coin randomized protocol that does not use many random coins. We refer the reader to [34, Chapter 3.3] for further details.

**Lemma 4.4** *Let $X = Y = \{0,1\}^n$ and $Z = \{0,1\}^m$. Any randomized protocol $\mathcal{P} : X \times Y \to Z$ may be transformed into a protocol $\mathcal{P}'$ so that (i) For every $x, y$ the statistical difference between $\mathcal{P}(x,y)$ and $\mathcal{P}'(x,y)$ is at most $\Delta$, (ii) $\mathcal{P}, \mathcal{P}'$ have the same communication complexity, and (iii) $\mathcal{P}'$ uses $O(\log n + \log \frac{1}{\Delta} + m)$ random bits.*

**Proof** It is sufficient to prove the claim for public-coin protocols. We will randomly transform $\mathcal{P}$ into $\mathcal{P}'$ so that with high probability the three requirements in the claim hold.

Denote by $\mathcal{P}_r(x,y)$ the (deterministic) outcome of protocol $\mathcal{P}$ when run with random coins $r$. Protocol $\mathcal{P}'$ is constructed by selecting at random $t$ strings $r_1, r_2, \ldots, r_t$. On input $(x,y)$ protocol $\mathcal{P}'$ selects $i \in_R \{1, \ldots, t\}$ and runs $\mathcal{P}_r(x,y)$.

Consider inputs $x, y$. Let $I(r,z)$ be 1 if $\mathcal{P}_r(x,y) = z$ and 0 otherwise. Let $\delta_z = \mathbf{E}[I(r,z)] = \Pr_r[\mathcal{P}(x,y) = z]$ be the probability that on input $(x,y)$ protocol $\mathcal{P}$ outputs $z$. Let $\delta'_z = \frac{1}{t}\sum_{i=1}^{t} I(r,z)$ be the probability that on input $(x,y)$ protocol $\mathcal{P}'$ outputs $z$. Note that $\mathbf{E}[\delta'_z] = \delta_z$. Setting $\varepsilon = \frac{\Delta}{2^m \delta_z}$ we get by Chernoff bounds that the statistical difference between $\mathcal{P}(x,y)$ and $\mathcal{P}'(x,y)$ is more than $\Delta$ with probability at most

$$\Pr_{r_1,r_2,\ldots,r_t}[\text{exists } z \text{ s.t. } |t\delta'_z - t\delta_z| > \varepsilon\delta t] \le \sum_z e^{-\varepsilon^2\delta_z t/3} < \sum_z 2^{-\Delta^2 t/(3\cdot 4^m \delta_z)} \le 2^m 2^{-\Delta^2 t/(3\cdot 2^m)}$$

The probability that there exist inputs $x, y$ for which the statistical difference is more than $\Delta$ is thus bounded by $2^{2n+m}2^{-\frac{\Delta^2 t}{3\cdot 2^m}}$ taking $t = O(\frac{2^m}{\Delta^2}(n+m))$ makes this probability exponentially small. Note that $\log t = O(\log n + \log \frac{1}{\Delta} + m)$ as required. $\square$

A straightforward way to use Lemma 4.4 is to take $\Delta$ to be negligible in the security parameter. It follows then that for every $x, y$ the outcomes of $\mathcal{P}'(x, y)$ and $\mathcal{P}(x, y)$ are statistically close (hence - computationally indistinguishable). Thus it suffices to show how to privately compute $\mathcal{P}'(x, y)$. This may be done by Alice and Bob choosing shares of the private and public coins used in the protocol and then applying Theorem 4.1 to the protocol that (i) retrieves the random coins from the shares, and (ii) simulates $\mathcal{P}'$. This transformation generally results in super-polynomial computational work.

## 4.5 Applications

For many interesting functions, there exists protocols with communication complexity that is significantly lower than the input size (and hence their circuit complexity). For such functions, there is a potential of getting protocols that are much more efficient than those resulting from the garbled circuit transformation. We now show how to apply the oblivious branching program representation to obtain good protocols for several such functions such as the Millionaires problem and position-wise inequality. We also consider efficient protocols for Karchmer-Wigderson games.

### 4.5.1 The Millionaires problem – who is larger?

Alice has input $x \in \{0, 1\}^n$ and Bob $y \in \{0, 1\}^n$ and they wish to decide whose input is larger. A circuit for deciding whether $x$ or $y$ is larger has $2n$ inputs and thus a protocol simulating it will perform $\Theta(n)$ invocations of the $\mathrm{OT}_1^2$ protocol. On the other hand it is known that the randomized communication complexity of this function is $O(\log n + \log 1/\varepsilon)$ where $\varepsilon$ is the probability of error. We describe a simple randomized protocol for the problem that results in an oblivious branching program with $O(\log(n) \cdot \log(\frac{1}{\varepsilon}))$ cost and linear (in $n$) width (it is possible to reduce it and obtain cost $O(\log n + \log \frac{1}{\varepsilon})$ as well).

Let $x_i$ (respectively, $y_i$) denote the $i$th bit of $x$ (resp. $y$) and let $\vec{x}_i = x_1, x_2, \ldots, x_{i-1}$ denote the $i$th prefix of $x$ (and resp. $\vec{y}_i$). The idea underlying all known communication efficient protocols for the problem is to perform a binary search for the position $i$ for which $x_i$ and $y_i$ differ but $\vec{x}_{i-1} = \vec{y}_{i-1}$.

To be able to execute the protocol Alice and Bob have to check whether $\vec{x}_i = \vec{y}_i$ or not. This is done by applying a hash function $h_i$ to $\vec{x}_i$ and $\vec{y}_i$ and then applying the branching program of Figure 3 on the result. The hash function $h_i : \{0, 1\}^i \mapsto \{0, 1\}^a$ is chosen so that for any two $\vec{x}_i \neq \vec{y}_i$ the probability that $h_i(\vec{x}_i) = h_i(\vec{y}_i)$ is smaller than $\frac{\varepsilon}{\log n}$ and hence $a = O(\log \frac{1}{\varepsilon})$ (we assume that $\varepsilon < \frac{1}{n}$ and in fact negligible in $n$). We would like to define $n$ such functions for all domains of size 2 through $2^n$ based on relatively few bits. This can be done in several ways, in particular, by sharing $O(\log \frac{1}{\varepsilon})$ random bits that induce $\frac{\varepsilon}{\log n}$-biased vectors, sampled via a random walk on an expander.

Given the choice of $\{h_i\}_{i=1}^n$ the communication protocol performs a binary search for the longest common prefix and then compares the next bit. The equality test for prefixes of length $i$ is done by Alice and Bob computing $h_i(\vec{x}_i)$ and $h_i(\vec{y}_i)$ respectively and then applying the branching program of Figure 3. The total error probability is $\varepsilon$, since there are $\log n$ steps. We now show the branching program complexity of implementing such a protocol.

Once Alice and Bob have agreed on $h_1, h_2, \ldots h_n$ they can define a branching program of width linear in $n$ to perform the binary search. Intuitively, all that the program needs to store is the current node in the binary search tree (this implies the index $i$ for which prefix equality should be decided) and the status of the string equality branching program (i.e. whether we have determined equality so far or not regarding prefix $i$). Therefore the total number of $\mathrm{OT}_1^{O(n)}$ performed is $a \log n$ which is $O(\log n \cdot \log \frac{1}{\varepsilon})$. A more careful implementation based on noisy binary search [15] can yield $O(\log n + \log \frac{1}{\varepsilon})$ many $\mathrm{OT}_1^{O(n)}$.

Note that these results mean that whether one is interested in a *communication* efficient protocol or a whether in a *computational* efficient protocol this is a good approach. The difference would be in the type of $OT$ used.

**Karchmer-Wigderson games** Let $f : \{0,1\}^n \to \{0,1\}$ be a Boolean function. In a Karchmer-Wigderson Alice's input satisfies $x \in f^{-1}(1)$ and Bob's input satisfies $y \in f^{-1}(0)$. The goal of the Karchmer-Wigderson game $G_f$ corresponding to $f$ is to find and index $i$ so that $x_i \neq y_i$. A related problem is the *universal relation* where Alice and Bob are required to find an index $i$ so that $x_i \neq y_i$, given that $x \neq y$. Note that the above protocol for the Millionaires problem actually finds the first index index $i$ where $x, y$ differ. Thus, its output may be modified to output $i$. Hence it serves as a communication efficient private protocol for the universal relation[16]. Clearly, a private protocol for the universal relation serves as a private protocol for any Karchmer-Wigderson game.

### 4.5.2 Position-wise inequality

Alice and Bob each have as inputs a list of $n$ elements $x_1, \ldots, x_n$ and $y_1, \ldots, y_n$ respectively, each element is in $\{0,1\}^n$. They want to determine whether there exists an index $i$ such that $x_i = y_i$ or not. This problem was suggested by Furer [18] as demonstrating a function with low Las Vegas communication complexity, compared with the deterministic one. Applying a private protocol for this task based on a circuit computing the relation results in communication and work corresponding to $\Omega(n^2)$ invocations of the $\mathrm{OT}_1^2$ protocol (due to the size of the inputs).

We present a branching program protocol for the problem with cost $O(n)$ and width poly$(n)$. Let $r_1, r_2, \ldots, r_m$ be $m > n$ random strings, each of length $n$. Let $\mathcal{X}_{i,j} = \langle x_i, r_j \rangle$ be the inner product of $x_i$ and $r_j$ and $\mathcal{Y}_{i,j} = \langle y_i, r_j \rangle$. Each layer of the branching program contains $mn$ states denoted $(i, j)$ for $1 \leq i \leq n$ and $1 \leq j \leq m$.

[16]An exact definition of private protocols for relations is out of the scope of this work.

**Initial State:** State $(1,1)$ in layer $L_0$.

**Transitions:** In state $(i,j)$ of layer $L_\ell$:
If $\mathcal{X}_{i,j} = \mathcal{Y}_{i,j}$ move to state $(i, \min(j+1, m))$ of layer $L_{\ell+1}$, otherwise move to state $(\min(i+1, n), \min(j+1, m))$ of layer $L_{\ell+1}$.

**Value:** In layer $L_m$ state $(i,m)$ has value 1, all other states have value 0.

It is easy to see that with $m = O(n)$ layers the probability that the branching program errs is negligible in $n$, hence we can apply Lemma 4.3 to get a private protocol.

# 5 Sublinear algorithms: Dealing with tables

The generalized indirect indexing primitive (recall Definition 3.2 and Construction 3.3 of $P_{GInd}$) underlies all our constructions so far, and hence these constructions make extensive use of *tables*. However, the usage was so far limited to *reading* from tables without *writing*. Furthermore, in the read pattern of $P_{GInd}$ is very specific – every table is accessed only once.

The limited usage of tables is reflected in $\mathcal{P}_{CC}^{PRIV}$ (Construction 4.1) and $\mathcal{P}_{BP}^{PRIV}$ (Construction 4.2) that use $P_{GInd}$ as a subroutine. In a typical usage of these protocols the parties create tables (as induced by their individual inputs) and then apply $\mathcal{P}_{CC}^{PRIV}$ or $\mathcal{P}_{BP}^{PRIV}$. However, once these tables are created, their values remain static.

**Overview of the section** In this section we consider the computational model of circuits with look-up tables (LUT), that allows a more elaborate usage of tables. We show that this model is strong enough to capture write-oblivious computations on a RAM machine. That is, computations that are allowed to perform any read operation, but whose write operations are restricted in the sense that their location and timing are not affected by the actual input. Further, we show that any computation on a RAM machine may be translated into a computation on a write-oblivious RAM machine with a modest blow-up, and hence, may be emulated by a circuit with LUT.

We start with a discussion of circuits with LUT and oblivious computation, stating informally some of the goals and results of this section. We then go into a more detailed account of our results. We begin by defining the LUT primitive and show how to evaluate it privately (using indirect indexing). The protocol for private LUT serves as a building block in a protocol for privately evaluating circuits with LUT – a variant of the garbled circuit transformation. We give several applications of circuits with LUT, resulting in efficient secure function evaluation. In particular, we show how to simulate any RAM machines in our model.

**Note 5.1** *We deal only with deterministic computation. Randomness may be handled similarly to Section 4.4.*

## 5.1 Circuits with LUT and write-oblivious RAM computation

We discuss our computational model - circuit with look-up tables – that allows an elaborate usage of tables. In this model, the gates of a circuit are look-up tables (LUT). The LUT input wires define the table entries and an index and the LUT output wires are set according to the value stored in the indexed position. We define a look-up table (LUT) primitive and show how to evaluate it privately (using indirect indexing). The protocol for private LUT serves as a building block in a protocol for privately evaluating circuits with LUT – a variant of the garbled circuit transformation.

Circuits with LUT amount to performing computations with tables as follows:

**Read operations:** The table values, as well as the index specifying the location of the read item are either preset or the result of an intermediate computation. In particular, it is possible to perform any kind of indirect read.

**Write operations:** The value written to the table may be the result of an intermediate operation but the location should be pre-determined. (In other words, no indirect writes are allowed.)

It follows that any computation on a RAM machine where write operations are *oblivious* – in the sense that the time and location of the write operations should not depend on the input and randomness – may be emulated by circuits with LUT.

### 5.1.1 Simulation by oblivious computation

An *oblivious* machine accesses memory locations in sequence that does not depend on the input. For example, an oblivious Turing Machine is a Turing machine for which the sequence of head positions (as a function of time) does not depend on the input. An oblivious RAM machine is a RAM machine is defined analogously. As discussed above, any computation on a RAM machine where write operations are oblivious may be emulated by circuits with LUT. Hence, we consider a computational model of a *write-oblivious* RAM machine, that may perform any read operation, but only oblivious write operations, in the sense that their time and location does not depend on the input. Below (Section 5.3) we show how to simulate any RAM machine computation an a write-oblivious RAM machine with a polylogarithmic blowup.

**Simulation by oblivious Turing Machines** The transformation of arbitrary machines into oblivious machines (computing the same function) was previously considered in the literature in different contexts. For Turing machines, Pippenger and Fischer [45] show that any computation on a (non-oblivious) one-tape Turing Machine may be simulated by a two-tape oblivious Turing Machine with a moderate cost. For any one-tape Turing Machine $\mathcal{M}$ computing $f_{\mathcal{M}}$ in time $O(T(n))$ there exists an oblivious two-tape Turing Machine computing $f_{\mathcal{M}}$ in time $O(T(n)\log T(n))$.

An important consequence of the result by Pippenger and Fischer [45] is that computation on a Turing Machine is not more efficient than circuits. For any Turing Machine $\mathcal{M}$ computing $f_{\mathcal{M}}$ in time $T(n)$ there exists a series of circuits $\{\mathcal{C}_n\}_{n\in\mathbb{N}}$ of size $|\mathcal{C}_n| = O(T(n)\log T(n))$ so that $\mathcal{C}_n$ computes $f_{\mathcal{M}}$ on inputs of length $n$. On the other hand, the oblivious simulation by probabilistic RAM does not seem to imply such a consequence on the relation between RAM machines and circuits. In particular, it is not known whether for any RAM machine $\mathcal{M}$ computing $f_{\mathcal{M}}$ in time $T(n)$ and space $S(n)$ there exists a series of circuits of size $o(T(n) \cdot S(n))$.

**Simulation by oblivious RAM** The access pattern of an oblivious machine does not depend on its input, hence it hides all information about the input (except its size). Goldreich and Ostrovsky [19, 24] utilized this property of oblivious machines to protect software from leaking its memory access sequence. However, to allow for efficient simulation of a RAM machine by oblivious RAM machine, they extend the definition of obliviousness to the probabilistic RAM machines (we emphasize that this extension of obliviousness differs from our extension of obliviousness). On such a machine, the probability distribution of the sequence of memory accesses should be independent of the machine input. They show that any RAM machine computing $f$ in time $T(n)$ with $s(n)$ memory locations can be simulated by an oblivious probabilistic RAM machine in time $O(T(n)(\log T(n))^3)$ with $O(s(\log s)^2)$ memory locations.

### 5.1.2 Closing the gap between circuits and RAM machines

The results of Pippenger and Ficher [45] imply that when considering circuits vs. Turing Machines, there is no significant advantage to the latter, since there exists a series of circuits of size comparable to the running time of the Turing Machine. A natural question is whether a similar result applies also for circuits vs. RAM machines. Currently, it is not known whether for every RAM machine there exists a series of circuits of size $o(T(n) \cdot S(n))$ computing the same function (where $T(n), S(n)$ denote the time and space used by the machine). Hence there is a potential gap – a computation on a RAM machine may be much more efficient than any circuit family.

We show that for circuits with LUT this gap is closed. Using the observation that for any write-oblivious RAM machine $M$ running in time $T(n)$ there exists a family of circuits with LUT of size $T(n)$ computing $f_{\mathcal{M}}$, all we need to show is an efficient simulation of any RAM machine via a write-oblivious RAM machine. We show such a simulation with a polylogarithmic blow-up. Hence, for a RAM machine $\mathcal{M}$ running in time $T(n)$ using space $S(n)$ there exists a series of circuits with LUT of size $T(n) \cdot \mathrm{polylog}(S(n))$ computing $f_{\mathcal{M}}$. Leading, potentially, to much more efficient secure function evaluation protocols.

Note that, unlike the transformation by Pippenger and Fischer [45], our simulation does not seem to imply the existence of small circuits (without LUT) for $f_{\mathcal{M}}$, since read operations are still non-oblivious.

## 5.2 Circuits with look-up tables

The look-up table (LUT) primitive retrieves an indexed table entry privately. On input, the table values as well and the index of the retrieved element are shared between the parties. On output, the retrieved entry is shared between the parties. Nothing is assumed about the way the LUT input values are computed and shared, but, the output shares must be random.

**Definition 5.1** PRIVATE ($LUT$) *Let* $\bar{R}_A = R_A[0], \ldots, R_A[w-1]$ *and* $\bar{R}_B = R_B[0], \ldots, R_B[w-1]$*. where* $R_A[i], R_B[i] \in \{0,1\}^m$*. The function* $LUT$ *is defined as follows:*

$$LUT((j_A, \bar{R}_A), (J_B, \bar{R}_B)) = (\pi, \pi \oplus R_A[j_A \oplus j_B] \oplus R_B[j_A \oplus j_B])$$

*where* $\pi \in_R \{0,1\}^m$*. The* $LUT$ width *is* $w$*.*

In the construction of $\mathcal{P}_{LUT}$ we use the indirect indexing primitive (Section 3). Let $j$ be an index shared by the parties. The protocol for privately computing $LUT$ uses $\mathcal{P}_{Ind_{AB}}$ so that the parties holds shares of $R_A[j]$ and $\mathcal{P}_{Ind_{BA}}$ so that they holds shares of $R_B[j]$. Then every party combines the shares he got. The cost of the resulting protocol is two $\mathrm{OT}_1^w$ invocations.

**Construction 5.1** (PRIVATE PROTOCOL FOR $LUT$) *Let* $\mathcal{P}_{LUT}^{\mathrm{PRIV}}((j_A, \bar{R}_A), (J_B, \bar{R}_B))$ *be the following protocol:*

**Inputs:** Alice and Bob share the index and the LUT values i.e. $j = j_A \oplus j_B$ and $R[i] = R_A[i] \oplus R_B[i]$ for $0 \leq i \leq w-1$.

**Step 1.1:** Alice and Bob run protocol $\mathcal{P}_{Ind_{AB}}(j_A, (j_B, R'_B))$ denote the outcome as $(s_A^1, s_B^1)$.

**Step 1.2:** Alice and Bob run protocol $\mathcal{P}_{Ind_{BA}}((j_A, R'_A), j_B)$ denote the outcome as $(s_A^2, s_B^2)$.

**Step 2:** Alice and Bob locally output $s_A^1 \oplus s_A^2$ and $s_B^1 \oplus s_B^2$ respectively.

**Note 5.2** *Steps 1.1 and 1.2 of* $\mathcal{P}_{LUT}^{\mathrm{PRIV}}$ *may be run in parallel, so that the round complexity of* $\mathcal{P}_{LUT}^{\mathrm{PRIV}}$ *equals that of* $\mathcal{P}_{Ind_{AB}}$*.*

Our definition of the outputs and inputs of the LUT primitive complies with the garbled circuit transformation in which the value of every wire is shared by the parties (see [20, 43]). Hence, it is possible to use the LUT primitive with the garbled circuit transformation as follows. The values on the LUT input wires define the table entry and an index; the LUT output wires are set the value stored in the indexed position.

Since it is possible to simulate any Boolean gate using LUT, every circuit may be simulated this way[17]. Moreover, a LUT may describe a complex gate with multiple inputs and outputs, with the cost of a single application of $\mathcal{P}_{LUT}$, hence the number of LUTs needed to construct a circuit for a function $f$ may be much smaller than the number of gates in the Boolean circuit for $f$, and the resulting computation and communication costs may be much lower. In particular, the communication of the resulting protocol may be sub-linear in the input size.

Using Construction 5.1 we get:

[17]However, the garbled circuit construction described in [43] usually results in a more efficient protocol.

**Claim 5.1** *Let $\mathcal{C}$ be a circuit with $s$ look-up tables of widths $w_1, \ldots, w_s$.*

*There exists a protocol privately computing $\mathcal{C}$ with computational work $O(\sum_{\ell=1}^{s} W_{\text{OT}}(w_\ell))$ and communication $O(\sum_{\ell=1}^{s} C_{\text{OT}}(w_\ell))$.*

### 5.2.1 Applications

We mention two types of applications: (i) functions that are computable by probing small portions of large data held by the parties, and (ii) function that are readily expressed using tables.

**Property testing** A great source of applications for circuits with LUT is property testing, where a (typically large) object is locally queried in order to check whether some global property holds [7, 47, 22]. I.e. the number of queries to the object is very small with respect to the object size (at the extreme - independent of the object size). The object is assumed either to have the property or to be far from any object having the property. Property testing algorithms are known to exists for a variety of properties including algebraic properties (e.g. being a polynomial of low degree) [7, 47] and graph properties (e.g. bipartiteness and $k$-colorability) [22].

Applying the garbled circuit to property testing algorithms results in a highly inefficient protocol, since the circuit size is at least the object size. It is, in principle, possible to directly apply Theorem 4.1 to get a communication efficient protocol, but the computational work done by the parties may get high (due to the large size of communication complexity trees involved).

On the other hand, expressing property testing algorithms as circuits with LUT results in very efficient protocols. The circuit consists of three types of components: (i) components which determine the next location to be probed in the object[18] (ii) components that probes the object (iii) a component that determines the answer. The component that probes the object is a circuit with LUT. The inputs for components of type (i) and (iii) are small (of order polylogarithmic in the input size, and polynomial in some 'security' parameter), and thus the circuits computing them are small. For these components, one may use for example the garbles circuit construction.

**Computing on encrypted data** Halevi and Mironov [25, 39] noted that some popular private-key encryption functions (e.g. *Data Encryption Standard* – DES and *Advanced Encryption Standard* – AES) are defined with look-up tables, hence they are readily expressible via circuits with LUT – suggesting a potentially efficient protocol.

Halevi suggested the problem of encrypting a message with a key shared by Alice and Bob. Alice holds $k_A$ and Bob holds $k_B$ and the outcome of the protocol is $\text{AES}_{k_A \oplus k_B}(m)$. A potential usage is for applications that use $\text{AES}_k(m_0)$ (with a fixed $m_0$) as a substitution for a 'random' function of $k$. Another usage is where both the message $m$ and the encryption key $k$ are shared between the parties, who wish to compute $\text{AES}_k(m)$.

Mironov suggested the problem of deciding, given an encryption, whether it is an encryption of a specific message without revealing the message, or the encryption key. In this setting, Alice holds an encryption key $k$ and Bob holds an encryption of a message $m$ (unknown to him) $\mathrm{DES}_k(m)$ and a 'candidate' message $m'$. At the end of the protocol Bob and Alice learn whether $m = m'$ or not.

[18] Sometimes this component reduces to choosing a random sample.

### 5.2.2 Sorting

We show how to construct a circuit with LUT that sorts an array of values. Our circuit implements a merge-sort algorithm. In Section 5.3 we use this sorter in our simulation of a general computation (on a RAM machine) by a computation that uses LUT (the more efficient simulation uses only the merger described below).

We describe the construction of the merger. The sorter is constructed in a standard way, applying the merger recursively. The input of the merger consists of two sorted arrays, each with $n$ elements $a[0], \ldots, a[n-1]$ and $b[0], \ldots, b[n-1]$. The output is an array $c[0], \ldots, c[2n-1]$ containing the merged lists.

The merger is depicted in Figure 4. It uses as a building block a circuit $\mathcal{C}$ for the task of updating the $j$th element of $c$. The inputs to $\mathcal{C}$ are the indices $i_a, i_b$ corresponding to the next element from arrays $a, b$ to be put into $c$. The outputs are $c[j]$ and updated indices $i_a, i_b$. $\mathcal{C}$ retrieves $a[i_a], b[i_b]$. If $i_a = n$ or $b[i_b] \leq a[i_a]$ the output is $b[i_b], i_a, i_b + 1$, otherwise it is $a[i_a], i_a + 1, i_b$.

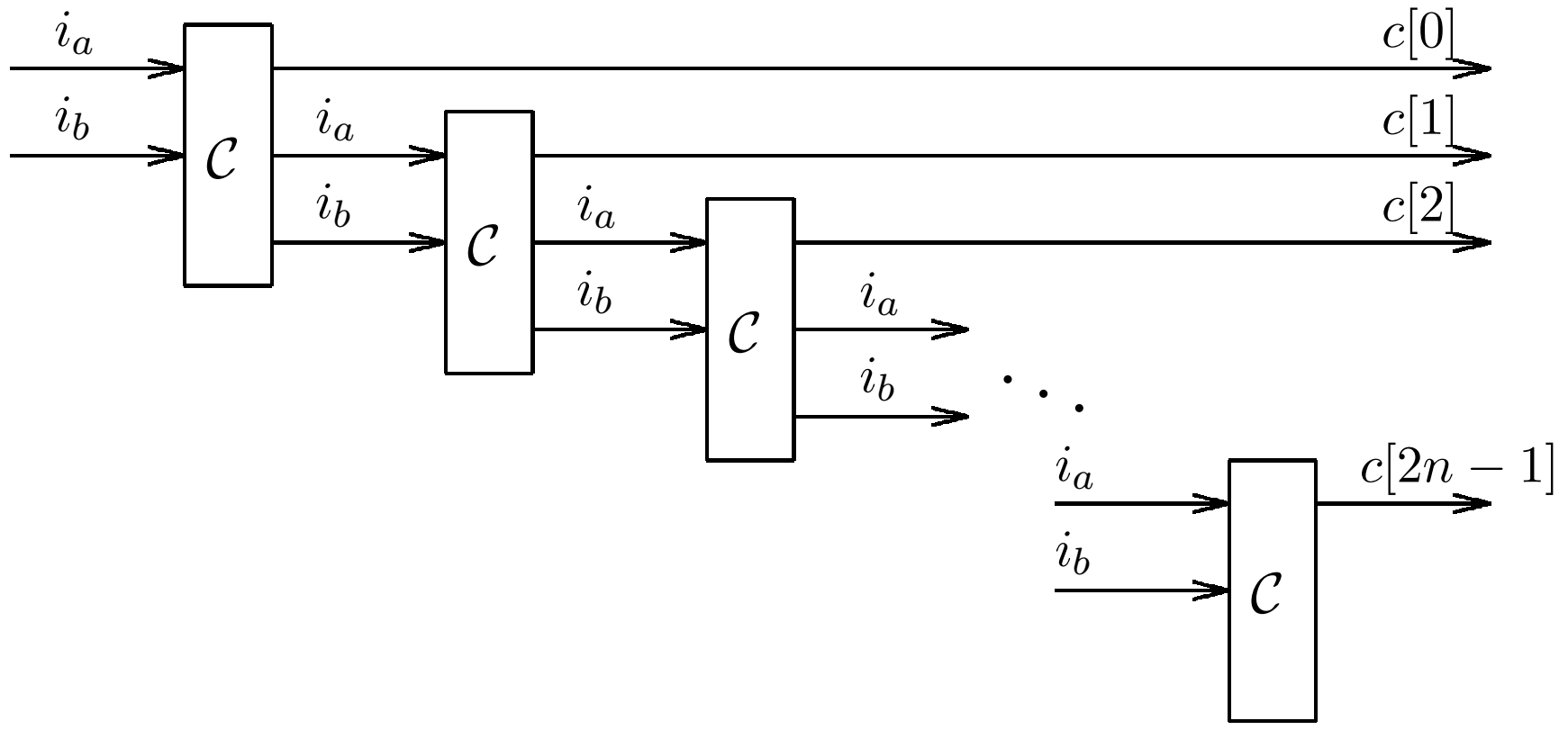

Figure 4: A merger

**Note 5.3** *An alternative to our sorter is a* sorting network. *I.e. a comparison network for which the output sequence is monotonically increasing for every input sequence. For instance one may use the AKS sorting network that achieves optimal size ($O(n \log n)$) and depth ($O(\log n)$) [2].*

*On the other hand, a* merging network *(i.e. a comparison network that merge two sorted inputs into one sorted output) is not as efficient as our merger, since any merging network requires $\Omega(n \log n)$ comparators.*

## 5.3 Simulating RAM by write-oblivious RAM

The RAM (random access memory) machine model is of a 'realistic' machine that can perform operations involving registers and main memory, where the memory may be indexed directly or indirectly both for read and write operations (we will defer from a specific definition of a RAM machine). A write-oblivious RAM machine is a RAM machine where the time and location of write operations depend solely on the input size (but not on the specific input or randomness).

Every computation on a write-oblivious RAM machine may be simulated by circuits with LUT in a step-by-step manner. A typical step (i) retrieves the current operation (indexed by some register) (ii) read from memory to registers (iii) manipulate registers (usually requires a small circuit), and (iv) write to memory. Each wire in the circuit with LUT simulating the computation corresponds to the a bit of memory or a register. For every write operation, the corresponding wire(s) are 'replaced' with the wires that carry the 'fresh' values.

**A basic simulation with $O(s)$ blowup** A straight forward simulation of RAM machines by write-oblivious RAM machines is to simulate every write operation by re-writing every memory location in a fixed (hence oblivious) order. To write value $\beta$ into address $\alpha$, Every memory location is read and then written with its 'current' value, unless when its address is $\alpha$, where it is written with value $\beta$. The cost hence is of $O(s)$ operations for every write operation that is simulated.

**Using logs** To save on the large overhead introduced by the basic simulation, we incorporate a scheme that does not apply it often. We postpone updates to the memory, and save them in a *log*. When many changes are accumulated we update the memory at once. Below, we describe such a scheme, that reduces the simulation blowup to $O(s^{1/2})$. Our final scheme maintains a hierarchy of logs, such that updates propagate through the logs. With this hierarchical construction we reduce the amortized cost of read and write operations to be polylogarithmic in $s$, resting in a polylogarithmic blowup. Our constructions are similar in nature to that of Goldreich and Ostrovsky [19, 24] and Ostrovsky and Shoup [44]. A similar luck of efficiency problem was addressed by Rosenblum and Ousterhout [46] in the context of file systems. They suggest log-structured file systems that optimize write operations to the file system by sequentially logging modifications to existing data rather than overwriting data in place.

### 5.3.1 A simulation with $O(s^{1/2})$ blowup

To simplify the presentation, we begin with an simulation that results in amortized cost of $O(s^{1/2})$ LUT accesses for every read/write operation. The main idea is to hold a log $L$ of $s^{1/2}$ elements that contains the most recent updates to the RAM memory. The entries of $L$ are address-value pairs $(\alpha, \beta)$ that reflect recent memory modifications. For every write operation a new entry is added to $L$ corresponding to the written value and its address. Once the log is 'full', its contents is sorted and then merged with that of the RAM. To merge the sorted table with the RAM we use a variant of the merger of Figure 4 that picks for every address $\alpha$ only its most recent pair.

**Read operations** To read an entry from address $\alpha$, the log $L$ is checked for the most recent address-value pair with address $\alpha$. If such a pair is found - the outcome of the read operation is set to be its value. Otherwise, the outcome is the LUT value at address $\alpha$.

**Write operations** To write $\beta$ into address $\alpha$, a new entry $(\alpha, \beta)$ is appended to the log $L$. Once the log $L$ contains $s^{1/2}$ entries (the log is 'full'), it is is sorted, merged with the RAM and reset to contain no elements.

The cost of a read operation is of $O(s^{1/2})$ accesses into $L$ and a single access to the LUT. The cost of a write operation is of a single access into $L$ (when $L$ is not full) and $O(s^{1/2} \log(s^{1/2})) + O(s) = O(s)$ LUT accesses once $L$ is full. The resulting amortized cost is of $O(s^{1/2})$ LUT accesses.

### 5.3.2 Achieving polylogarithmic blowup

We now generalize the above construction to achieve lower amortized costs. Instead of having a single log, we will maintain logs $L_0, L_1, \ldots, L_k$ of sizes $n_0, \ldots, n_k$ where $n_i = s/(\log s)^i$ and $k = O(\log s/\log\log s)$ (so that $n_k = 1$). On a write operation, a new address-value pair is entered into $L_k$. This pair gradually propagates through the logs using the rule that whenever the log $L_i$ is 'full' its contents is merged with that of log $L_{i-1}$. The address-value entries of every log are kept sorted according to the address. For simplicity, we assume that log $L_0$ is initialized to contain the $s$ pairs $(0,0), (1,0), \ldots, (s-1, 0)$. All other logs are initialized to be empty.

**Read operations** To read an entry from address $\alpha$ it is searched in $L_k, L_{k-1}, \ldots, L_0$ (since the tables are sorted, a binary search is possible). The outcome of the read operation is set to the value of the first matching address-value pair.

**Write operations** To write value $\beta$ into address $\alpha$, a new entry $(\alpha, \beta)$ is appended to log $L_k$. Let $i = k$. The following process is repeated as long as $i > 0$ and the number of elements in $L_i$ equals $n_i$:

- Merge log $L_i$ with $L_{i-1}$. Reset log $L_i$ to empty.
- Decrement $i$.

The cost of a read operation is of $O(\sum_{i=0}^{k} \log n_i) = O(\log^2 s/\log\log s)$ LUT accesses. The resulting amortized cost of a write operation is of $O(\sum_{i=0}^{k} n_i/n_{i+1}) = O(\log^2 s/\log\log s)$ LUT accesses.

**Using the transformation from RAM machine to obtain efficient secure function evaluation protocols** The transformation yield efficient protocols for functions that access (and possibly modify) only a small portion of their memory. This is similar to what we described regarding property checking, but more general.

Functions $f$ that may be expressed as the composition of local computations (by the parties over their inputs) with a global computation also yield efficient protocols. Suppose the computed function $f(x, y)$ can be expressed as the composition $h(g_A(x), g_B(y))$ where $g_A$ and $g_B$ are efficiently computable functions[19] and $h$ is sublinear in the write-oblivious RAM machine model. We get a secure function evaluation protocol for $f$ with sublinear communication.

# 6 Handling Malicious Parties

So far we have assumed that the parties are semi-honest. It is known that there is a general compiler – which is *not* communication preserving – from a protocol for semi-honest participants to malicious ones [23] (we refer to the description in [20]). This transformation *forces* the parties to act semi-honestly. In this section we briefly outline a transformation – from the semi-honest case to the malicious one – that is communication preserving in the sense that the communication is increased by a factor polynomially related to the security parameter.

Recall that the compiler has three components: *input-commitment*, *coin-generation* and *single step emulation*. Our transformation follows the same lines. Each party commits to its input and random bits. Once this is done then all that remains is for each party to show that it is following the protocol as specified for the semi-honest case. This means that following each step in the original protocol the sender in this step has to prove the consistency of the message with the input and random bits he is committed to.

However, the main difference with the transformation of [20] is that since the number of transferred bits should be at most polylog in the input size (and hence much smaller) the commitment to the input and randomness must be *computational*. Thus, there exist *many* possible valid opening following the commit step. It might even be the extreme case that for any message the sender gives there exists a proof of consistency with the original commitment. The security of the scheme should thus rest on the hardness of *finding* more than one valid opening for the commitment. To make the proof meaningful, the sender provides a zero-knowledge *proof of knowledge* that it knows an opening of the initial commitment that is consistent with the current message it has sent. Assuming the sender may know at most *one* such opening (or it has broken the commitment scheme) there is only *one* possible message in the compiled protocol for which he can provide such a proof. Since a proof of knowledge allows the simulator to extract the witness from the prover, it follows that if it is possible to extract two different assignments to the initial commitment then the commitment scheme is not binding.

## 6.1 Input commitment

The cryptographic primitive we need is a communication efficient commitment to a large string. Such a commitment scheme is used in Kilian's construction of zero-knowledge arguments [30]. It is based on any perfectly binding commitment scheme (applied to each bit of the committed string) combined with a hash tree of a collision intractable hash function. The communication complexity of the resulting input commitment scheme is thus poly($k$).

[19] It is possible that $g_A(x), g_B$ depend on common randomness, as described in Section 4.4.1.

## 6.2 Random coins generation

The (pseudo) random coins used by each party are the result of applying a pseudo-random generator $G$ to a seed of length $k$. Thus it is possible to use the random coins generation protocol as described in [20].

## 6.3 Zero knowledge arguments of knowledge of NP witnesses

The primitive we are after is a system of zero knowledge arguments of knowledge of NP witnesses that operates with communication complexity $\text{polylog}(n) \cdot \text{poly}(k)$. This problem has not been explicitly treated in the past. We observe however that the construction of zero knowledge arguments for NP by Kilian [30] (see also [31]), combined with the PCP system of Arora et al. [3] gives the desired properties.

Kilian's construction is based on the PCP Theorem [3] and uses a commitment scheme that allows to open the commitment for single bits. The prover is required to commit to a proof tape $\tau$ so that every bit $\sigma$ of $\tau$ is represented by the two bits $\sigma', \sigma''$ (called 'blob') randomly chosen so that $\sigma = \sigma' \oplus \sigma''$. The verifier simulates a PCP verifier and asks the prover to prove (in zero knowledge) the value of a predicate concerning $O(1)$ of the blobs. In this proof the verifier asks to open, at random, exactly one bit of these $O(1)$ blobs. It follows that if the prover succeeds in convincing the verifier with high probability then it is possible to extract a proof tape $\hat{\tau}$ that is accepted by the PCP verifier with high probability.

**Proposition 6.1** *The zero knowledge argument system of Kilian is a zero knowledge argument of knowledge for a PCP proof $\hat{\tau}$ that is accepted by the PCP verifier with probability at least $1 - \epsilon$.*

Given a PCP proof $\hat{\tau}$ that is accepted with probability $1 - \epsilon$ Our next step is to convert it into a proof $\tilde{\tau}$ that is accepted with probability 1. For this end, we consider the specific PCP system by Arora et al. [3]. The conversion from $\hat{\tau}$ to $\tilde{\tau}$ exploits the fact that all proofs that are accepted with high probability are close (in Hamming distance) to valid code-words of a linear error correcting code, so that they may be uniquely decoded to a valid codeword.

**Proposition 6.2 ([33] Theorem 4)** *For the PCP system of Arora et al., there exists an efficient procedure so that given a PCP proof $\hat{\tau}$ that is accepted by the PCP verifier with probability $1 - \epsilon$ outputs a PCP proof $\tilde{\tau}$ that is accepted by the PCP verifier with probability 1.*

Since an assignment for the original SAT formula is explicit in the PCP proof in the construction of Arora et al., we get as a consequence of propositions 6.1, 6.2 a communication preserving zero knowledge argument system of knowledge of NP witnesses.

# 7 Discussion and Open Problems

**Communication and computation** We have seen how to obtain secure protocols with reasonable computational requirements provided the parties are computable by polynomial width branching programs. One interesting issue this work raises is whether it is possible to start with a (general) communication complexity protocol and preserve *simultaneously* computation and communication[20].

One consequence of a transformation preserving both computation and communication is the non-existence of incompressible functions. Informally, a function $f$ is incompressible if to communicate $f(x)$ with communication complexity significantly smaller than $|f(x)|$ one has send $x$. I.e. there is no way to send $f(x)$ efficiently without revealing $x$. We refer the reader to [13] for definition and applications of incompressible functions.

To understand the power of our technique we should investigate the class $CCW(c, w)$ - functions computable by a branching program protocols of width $w$ and communication cost $c$. For functions $f \in CCW(o(n), \text{poly}(n))$ Theorem 4.2 yields secure protocols with sub-linear OT invocations and polynomial work. Here the results of Beame et al. [10] may be relevant, but their notion of size is stronger than we need.

**The multi-party setting** The method of translating protocols in the communication complexity model and branching program protocols into private protocols using indirect indexing extends for a class of protocols with more than two parties. These protocols are *oblivious* in the sense that the communication *pattern* (i.e. which party sends a message at which stage) does not depends on the input.

**Efficient transformation for malicious parties** The compiler described in [20] as well as the adjustment we describe in Section 6 is inefficient in terms of computational work. The overhead incurred in applying Cook's Theorem (to translate an NP statement to SAT) and then the PCP Theorem is prohibitive in many applications. An approach that might prove useful is to have a transformation for protocols in the semi-honest model with specific properties, e.g. for protocols in which the parties do not learn anything (even if malicious) until the very last step.

# Acknowledgments

We thank Madhu Sudan for a discussion of the reversibility of PCP constructions and pointing [33], Yehuda Lindell for discussions regarding handling malicious parties, and Yuval Ishai for various comments and discussions.

[20] In Appendix A we give a variant of the garbled circuit transformation with reduced work.

# A A variant on the garbled circuit transformation

In the following we give a variant on the garbled circuit construction that allows for less computational work. Specifically, we show how to apply the garbled circuit construction to an (insecure) protocol for computing $f$ so as to reduce the use of the oblivious transfer primitive. The number of oblivious transfers invocations in our construction equals the communication complexity of the original protocol computing $f$. The communication is proportional to the sizes of the circuits involved in the computation.

Consider an (possibly insecure) protocol for a functionality $f$. The protocol is defined by two circuits – one circuit computes Alice's next message given her input, her randomness and the messages she received so far, the other circuit computes Bob's next message. For simplicity, we assume that Alice and Bob alternately send single bit messages (Alice sends the first message), resulting in the computation of $f$ by Alice. We denote by $c$ the number of messages sent in the protocol (without loss of generality, we assume $c$ to be even).

Note that once Alice sees her input and chooses her randomness she may 'hard-wire' them into her circuit, resulting in a circuit $\mathcal{C}_{Alice}$. The only inputs of $\mathcal{C}_{Alice}$ are the messages received from Bob. Similarly, Bob may hard-wire his input and randomness into his circuit, resulting in $\mathcal{C}_{Bob}$.

**Protocol representation** In the following we slightly abuse the term *circuit* and allow for circuits with multiple input (and outputs), so that the inputs are not given at once. Given a subset of the inputs, the circuit computes the outputs of the sub-circuit depending solely on these inputs.

Specifically, we denote the inputs of Alice's circuit $\mathcal{C}_{Alice}$ by $m_1, m_3, \ldots, m_{c-1}$ (corresponding to the messages she receives from Bob) and its outputs by $m_0, m_2, \ldots, m_{c-2}$ and $z$ (corresponding to the messages she sends and the outcome of the computation). The output $m_\ell$ is computed after inputs $m_1, m_3, \ldots, m_{\ell-1}$ are specified; $z$ is computed after all the inputs are specified. Similarly, the inputs to Bob's circuit $\mathcal{C}_{Bob}$ are denoted by $m_0, m_2, \ldots, m_{c-2}$, and the outputs by $m_1, m_3, \ldots, m_{c-1}$. Output $m_\ell$ is computed by $\mathcal{C}_{Bob}$ after inputs $m_0, m_2, \ldots, m_{\ell-1}$ are specified.

We now show how to modify the garbled circuit construction, as described in [43], so as to achieve our protocol. There are two main differences between our construction below and that described in [43]:

1. In [43] one party holds the input and the other holds a circuit. The protocol below is symmetric in the sense that both parties hold circuits and control the inputs to these circuits.

2. The protocol in [43] runs in a constant number of rounds. The round complexity of the protocol below is a function of $c$, the number of messages sent in the original protocol. Correspondingly, the inputs to the circuits are not given at once, but are generated during the run of the protocol.

## A.1 A protocol for computing $f(x, y)$

Alice's input to the protocol is $x$, Bob's input is $y$. We assume that Alice and Bob agree on circuits that, given their inputs and randomness, determine the messages sent in the protocol. Alice and Bob hard-wire their inputs and randomness in these circuits, resulting in $\mathcal{C}_{Alice}$ and $\mathcal{C}_{Bob}$. These are the inputs to the garbling procedure below. Without loss of generality, we assume a disjoint numbering on the wires of $C_{Alice}$ and $C_{Bob}$. Our description follows closely the notation of [43].

**Garbling the circuits:** Alice assigns for each wire $i$ of $\mathcal{C}_{Alice}$ a random bit $\pi_i \in_R \{0, 1\}$ and two random $k$-bit values $(W_i^0, W_i^1)$ corresponding to 0/1 values of the wire. For a wire $i$ with value $b_i$, its 'garbled' value is $\langle W_i^{b_i}, c_i \rangle$ where $c_i = \pi_i \oplus b_i$.

For every Boolean gate $g$ in $\mathcal{C}_{Alice}$ computing $b_k = g(b_i, b_j)$ Alice prepares a table $T_g$ with entries

$$c_i, c_j :< W_k^{g(b_i,b_j)}, c_k > \oplus F_{W_i^{b_i}}(c_j) \oplus F_{W_j^{b_j}}(c_i)$$

where $F$ is a pseudo-random function with output length $k + 1$. The table encrypts the garbled value of wire $k$ using the output of $F$ keyed by the garbled values of the input wires. Alice sends $T_g$ to Bob.

Similarly, Bob garbles his circuit and sends the corresponding tables to Alice.

**Executing the protocol:** The tables described above allow each party to compute the garbled output of every gate given its garbled inputs. We now describe how to compute the garbled inputs to the circuits.

Recall that very round of the protocol correspond to a message sent in the original protocol. We describe how to construct the input for round $\ell$. We consider the case where $\ell$ is even i.e. $m_\ell$ is a message sent from Alice to Bob (the odd $\ell$ case is treated symmetrically). Let $i$ be the (output) wire corresponding to $m_{\ell-1}$ in $\mathcal{C}_{Bob}$ and $j$ be the (input) wire corresponding to $m_{\ell-1}$ in $\mathcal{C}_{Alice}$. Assume that Alice completed the computation of (a garbled version of) Bob's message $m_{\ell-1}$ (this is the output of round $\ell - 1$). Namely, Alice holds $\langle W_i^{m_{\ell-1}}, c_i \rangle$ where $c_i = \pi_i \oplus m_{\ell-1}$ and $\pi_i$ is known to Bob.

The garbled value $\langle W_i^{m_{\ell-1}}, c_i \rangle$ may not be used in the evaluation of $\mathcal{C}_{Alice}$ since the garbling of wire $i$ is not related with that of wire $j$ (furthermore, had Bob learned $c_i$ he could compute $m_{\ell-1}$ and thus compromise the privacy of the protocol). Thus, Alice and Bob run the following procedure for translating this garbled value:

Alice creates a table corresponding to the two possible values of $\pi_i$ with entry $c_i$ being $\langle W_j^0, \pi_j \rangle$ and entry $1 - c_i$ being $\langle W_j^1, 1 - \pi_j \rangle$. Alice and Bob invoke a $\mathrm{OT}_1^2$ protocol so that Bob retrieves entry $\pi_i$ of the table. Since this entry is $\langle W_j^{m_{\ell-1}}, m_{\ell-1} \oplus \pi_j \rangle$ it is the garbled value of wire $j$ and hence Bob may now compute the garbled $m_\ell$.

Let $s$ be the total number of gates in $\mathcal{C}_{Alice}$ and $\mathcal{C}_{Bob}$. We get the following claim:

**Claim A.1** *There exists a secure function evaluation protocol for $f$ with computational work $c \cdot W_{OT}(2, k) + s \cdot W_{PRF}(k)$ and communication $c \cdot C_{OT}(2, k) + O(sk)$.*

Typically it is the case where $W_{PRF}$ is much smaller than $W_{OT}$. Thus, if the communication complexity of $f$ is much smaller than being linear ($c = o(n)$), Claim A.1 results in protocols which are much more efficient (in terms of computational work) than those resulting from the garbled circuit transformation described in [43] (and [20]).